\documentclass[12pt]{iopart}

\usepackage[pdftex]{graphicx}
\usepackage{color}
\usepackage{times} \normalfont \usepackage[T1]{fontenc}

\def\qe{\textsc{Quantum ESPRESSO}}

\begin{document}

\title[\qe]{\qe: a modular and open-source software project for quantum
  simulations of materials
}

\author{
  Paolo Giannozzi,$^{1,2}$
  Stefano Baroni,$^{1,3}$
  Nicola Bonini,$^4$
  Matteo Calandra,$^5$
  Roberto Car,$^6$
  Carlo Cavazzoni,$^{7,8}$
  Davide Ceresoli,$^4$
  Guido L. Chiarotti,$^9$
  Matteo Cococcioni,$^{10}$
  Ismaila Dabo,$^{11}$
  Andrea Dal Corso,$^{1,3}$
  Stefano Fabris,$^{1,3}$
  Guido Fratesi,$^{12}$
  Stefano de Gironcoli,$^{1,3}$
  Ralph Gebauer,$^{1,13}$
  Uwe Gerstmann,$^{14}$
  Christos Gougoussis,$^5$
  Anton Kokalj,$^{1,15}$
  Michele Lazzeri,$^5$
  Layla Martin-Samos,$^1$
  Nicola Marzari,$^{1,3}$
  Francesco Mauri,$^5$
  Riccardo Mazzarello,$^{16}$
  Stefano Paolini,$^{3,9}$
  Alfredo Pasquarello,$^{17}$
  Lorenzo Paulatto,$^{1,3}$
  Carlo Sbraccia,$^{1}$\footnote{Present address: 
    Constellation Energy Commodities Group 7th Floor, 61 Aldwich, London, 
    WC2B 4AE, United Kingdom}
  Sandro Scandolo,$^{1,13}$
  Gabriele Sclauzero,$^{1,3}$
  Ari P. Seitsonen,$^5$
  Alexander Smogunov,$^{13}$
  Paolo Umari,$^1$
  Renata M. Wentzcovitch,$^{18,19}$
}

\address{
  $^1$CNR-INFM Democritos National Simulation Center, 34100 Trieste, 
  Italy
} 

\address{
  $^2$Dipartimento di Fisica, Universit\`a degli Studi di Udine,
  via delle Scienze 208, 33100 Udine, Italy
}

\address{
  $^3$SISSA -- Scuola Internazionale Superiore di Studi Avanzati, 
  via Beirut 2-4, 34151 Trieste Grignano, Italy
}

\address{
  $^4$Department of Materials Science and Engineering,
  Massachusetts Institute of Technology, Cambridge, MA 02139 USA
}

\address{
  $^5$Institut de Min\'eralogie et de Physique des Milieux 
  Condens\'es, Universit\'e Pierre et Marie Curie, CNRS, IPGP,
  140 rue de Lourmel, 75015 Paris, France
}

\address{
  $^6$Department of Chemistry, Princeton University, Princeton NJ 08544, USA
}

\address{
  $^7$CINECA National Supercomputing Center, Casalecchio di Reno,
  40033 Bologna, Italy
}

\address{
  $^8$CNR-INFM S3 Research Center, 41100 Modena, Italy
}

\address{
  $^9$SPIN s.r.l. via del Follatoio 12, 34148 Trieste, Italy
}

\address{
  $^{10}$Department of Chemical Engineering and Materials Science,
  University of Minnesota, 151 Amundson Hall, 421 Washington Avenue 
  SE, Minneapolis MN 55455, USA
}

\address{
  $^{11}$Universit\'e Paris-Est, CERMICS, Projet Micmac ENPC-INRIA, 
  6-8 avenue Blaise Pascal, 77455 Marne-la-Vall\'ee Cedex 2, France
}

\address{
  $^{12}$Dipartimento di Scienza dei Materiali, Universit\`a degli Studi di 
  Milano-Bicocca, via Cozzi 53, 20125 Milano, Italy
} 

\address{
  $^{13}$The Abdus Salam International Centre for Theoretical Physics,
  Strada Costiera 11, 34151 Trieste Grignano, Italy
}

\address{
  $^{14}$Theoretische Physik, Universit\"at Paderborn, D-33098 Paderborn, Germany
}

\address{
  $^{15}$Jo\v zef Stefan Institute, Jamova 39, SI-1000 Ljubljana,
  Slovenia
}

\address{
  $^{16}$Computational Science, Department of Chemistry and Applied Biosciences,
  ETH Zurich, USI Campus, via Giuseppe Buffi 13, CH-6900 Lugano,
  Switzerland
}

\address{
  $^{17}$Ecole Polytechnique F\'ed\'erale de Lausanne (EPFL), Institute of 
  Theoretical Physics, and
  Institut Romand de Recherche Num\'erique en Physique des
  Mat\'eriaux (IRRMA), CH-1015 Lausanne, Switzerland
}

\address{
  $^{18}$Department of Chemical Engineering and Materials Science,
  University of Minnesota, 151 Amundson Hall, 421 Washington Avenue 
  SE, Minneapolis MN 55455, USA
}

\address{
  $^{19}$Minnesota Supercomputing Institute for Advanced Computational Research,
  University of Minnesota, Minneapolis MN 55455, USA
}

% \ead{paolo.giannozzi@uniud.it}

\begin{abstract}
  \qe\ is an integrated suite of computer codes for
  electronic-structure calculations and materials modeling, based on
  density-functional theory, plane waves, and pseudopotentials
  (norm-conserving, ultrasoft, and projector-augmented wave). \qe\
  stands for {\em opEn Source Package for Research in Electronic
    Structure, Simulation, and Optimization}. It is freely available
  to researchers around the world under the terms of the GNU General
  Public License. \qe\ builds upon newly-restructured
  electronic-structure codes that have been developed and tested by
  some of the original authors of novel electronic-structure
  algorithms and applied in the last twenty years by some of the
  leading materials modeling groups worldwide. Innovation and
  efficiency are still its main focus, with special attention paid to
  massively-parallel architectures, and a great effort being devoted
  to user friendliness. \qe\ is evolving towards a distribution of
  independent and inter-operable codes in the spirit of an open-source
  project, where researchers active in the field of
  electronic-structure calculations are encouraged to participate in
  the project by contributing their own codes or by implementing their
  own ideas into existing codes.
\end{abstract}

\maketitle

\section{Introduction}

The combination of methodological and algorithmic innovations and
ever-increasing computer power is delivering a {\em simulation
  revolution} in materials modeling, starting from the nanoscale up to
bulk materials and devices \cite{marzari-2006}.  Electronic-structure
simulations based on density-functional theory (DFT)
\cite{dft1,dft2,martin-book} have been instrumental to this
revolution, and their application has now spread outside a restricted
core of researchers in condensed-matter theory and quantum chemistry,
involving a vast community of end users with very diverse scientific
backgrounds and research interests. Sustaining this revolution and
extending its beneficial effects to the many fields of science and
technology that can capitalize on it represents a multifold
challenge. In our view it is also a most urgent, fascinating and
fruitful endeavor, able to deliver new forms for scientific
exploration and discovery, where a very complex infrastructure---made
of software rather than hardware---can be made available to any
researcher, and whose capabilities continue to increase thanks to the
methodological innovations and computing power scalability alluded to
above.

Over the past few decades, innovation in materials simulation and
modeling has resulted from the concerted efforts of many individuals
and groups worldwide, often of small size. Their success has been made
possible by a combination of competences, ranging from the ability to
address meaningful and challenging problems, to a rigorous insight
into theoretical methods, ending with a marked sensibility to matters
of numerical accuracy and algorithmic efficiency. The readiness to
implement new algorithms that utilize novel ideas requires total
control over the software being used--for this reason, the physics
community has long relied on in-house computer codes to develop and
implement new ideas and algorithms. Transitioning these development
codes to production tools is nevertheless essential, both to
extensively validate new methods and to speed up their acceptance by
the scientific community. At the same time, the dissemination of codes
has to be substantial, to justify the learning efforts of PhD students
and young postdocs who would soon be confronted with the necessity of
deploying their competences in different research groups.  In order to
sustain innovation in numerical simulation, we believe there should be
little, if any, distinction between development and production codes;
computer codes should be easy to maintain, to understand by different
generations of young researchers, to modify, and extend; they should
be easy to use by the layman, as well as general and flexible enough
to be enticing for a vast and diverse community of end users. One
easily understands that such conflicting requirements can only be
tempered, if anything, within organized and {\em modular} software
projects.

Software modularity also comes as a necessity when complex problems in
complex materials need to be tackled with an array of different
methods and techniques. Multiscale approaches, in particular, strive
to combine methods with different accuracy and scope to describe
different parts of a complex system, or phenomena occurring at
different time and/or length scales.  Such approaches will require
software packages that can perform different kinds of computations on
different aspects of the same problem and/or different portions of the
same system, and that allow for interoperability or joint usage of the
different modules.  Different packages should at the very least share
the same input/output data formats; ideally they should also share a
number of mathematical and application libraries, as well as the
internal representation of some of the data structures on which they
operate.  Individual researchers or research groups find it
increasingly difficult to meet all these requirements and to continue
to develop and maintain in-house software project of increasing
complexity.  Thus, different and possibly collaborative solutions
should be sought.

A successful example comes from the software for simulations in
quantum chemistry, that has often (but not always) evolved towards
{\em commercialization}: the development and maintenance of most
well-known packages is devolved to non-profit
\cite{adf,crystal,molcas,turbomole} or commercial
\cite{g03,ms,schrodinger,wavefunction} companies.  The software is
released for a fee under some proprietary license that may impose
several restrictions to the availability of sources (computer code in
a high-level language) and to what can be done with the software.
This model has worked well, and is also used by some of the leading
development groups in the condensed-matter electronic-structure
community \cite{vasp,castep}, while some proprietary projects allow
for some free academic usage of their products
\cite{cpmd,nwchem,nwchem2,gamess1,gamess2,castep}.  A commercial
endeavor also brings the distinctive advantage of a professional
approach to software development, maintenance, documentation, and
support.

We believe however that a more interesting and fruitful alternative
can be pursued, and one that is closer to the spirit of science and
scientific endeavor, modeled on the experience of the open-source
software community. Under this model, a large community of users has
full access to the source code and the development material, under the
coordination of a smaller group of core developers.  In the long term,
and in the absence of entrenched monopolies, this strategy could be
more effective in providing good software solutions and in nurturing a
community engaged in providing those solutions, as compared to the
proprietary software strategy.  In the case of software for scientific
usage, such an approach has the additional, and by no means minor,
advantage to be in line with the tradition and best practice of
science, that require reproducibility of other people's results, and
where collaboration is no less important than competition.

In this paper we will shortly describe our answer to the
above-mentioned problems, as embodied in our \qe\ project (indeed,
\textsc{ESPRESSO} stands for {\em opEn Source Package for Research in Electronic
  Structure, Simulation, and Optimization}).  First, in Sec. 2, we
describe the guiding lines of our effort. In Sec. 3, we give an
overview of the current capabilities of \qe. In Sec. 4, we provide a
short description of each software component presently distributed
within \qe. In Sec. 5 we give an overview of the parallelization
strategies followed and implemented in \qe.
Finally, Sec. 6 describes current developments and offers
a perspective outlook. The Appendix sections discuss some of the more
specific technical details of the algorithms used, that have not been
documented elsewhere.

\section{The \qe\ project}

\qe\ is an integrated suite of computer codes for electronic-structure
calculations and materials modeling based on density-functional
theory, plane waves basis sets and pseudopotentials to represent
electron-ion interactions.  \qe\ is {\em free}, open-source software
distributed under the terms of the GNU General Public License (GPL)
\cite{gpl}.

The two main goals of this project are to foster methodological
innovation in the field of electronic-structure simulations and to
provide a wide and diverse community of end users with highly
efficient, robust, and user-friendly software implementing the most
recent innovations in this field.  Other open-source projects
\cite{abinit,cp2k,dacapo,gpaw,PSI3} exist, besides \qe, that address
electronic-structure calculations and various materials simulation
techniques based on them.  Unlike some of these projects, \qe\ does
not aim at providing a single monolithic code able to perform several
different tasks by specifying different input data to a same
executable. Our general philosophy is rather that of an {\em open
  distribution}, i.e. an integrated suite of codes designed to be
interoperable, much in the spirit of a Linux distribution, and thus
built around a number of {\em core components} designed and maintained
by a small group of core developers, plus a number of
auxiliary/complementary codes designed, implemented, and maintained by
members of a wider community of users. The distribution can even be
redundant, with different applications addressing the same problem in
different ways; at the end, the sole requirements that \qe\ components
must fulfill are that: {\em i)} they are distributed under the same
GPL license agreement \cite{gpl} as the other \qe\ components; {\em ii})
they are fully interoperable with the other components. Of course,
they need to be scientifically sound, verified and validated.
External contributors are encouraged to join the \qe\ project, if they
wish, while maintaining their own individual distribution and
advertisement mode for their software (for instance, by maintaining
individual web sites with their own brand names \cite{wannier90}).  To
facilitate this, a web service called \texttt{qe-forge}
\cite{qe-forge}, described in the next subsection, has been recently
put in place.

Interoperability of different components within \qe\ is granted by the
use of common formats for the input, output, and work files. In
addition, external contributors are encouraged, but not by any means
forced, to use the many numerical and application libraries on which
the core components are built. Of course, this general philosophy must
be seen more as an objective to which a very complex software project
tends, rather than a starting point.

One of the main concerns that motivated the birth of the \qe\ project
is high performance, both in serial and in parallel execution.  High
serial performance across different architectures is achieved by the
systematic use of standardized mathematical libraries (BLAS, LAPACK
\cite{laug}, and FFTW \cite{FFTW05}) for which highly optimized
implementations exist on many platforms; when proprietary
optimizations of these libraries are not available, the user can
compile the library sources distributed with \qe.  Optimal performance
in parallel execution is achieved through the design of several
parallelization levels, using sophisticated communication algorithms,
whose implementation often does not need to concern the developer,
being embedded and concealed in appropriate software layers. As a result the
performance of the key engines, \texttt{PWscf} (Sec. \ref{sec:PWscf}) 
and \texttt{CP} (Sec. \ref{sec:CP}) may scale on massively parallel 
computers up to thousands of processors.

The distribution is organized into a basic set of modules, libraries,
installation utilities, plus a number of directories, each containing
one or more executables, performing specific tasks. The communications
between the different executables take place via data files. We think
that this kind of approach lowers the {\em learning barrier} for those
who wish to contribute to the project.  The codes distributed with
\qe, including many auxiliary codes for the post-processing of the data
generated by the simulations, are easy to install and to use. The GNU
\texttt{configure} and \texttt{make} utilities ensure a
straightforward installation on many different machines.  Applications
are run through text input files based on Fortran namelists, that
require the users to specify only an essential but small subset of the
many control variables available; a specialized graphical user
interface (GUI) that is provided with the distribution facilitates
this task for most component programs.  It is foreseen that in the
near future the design of special APIs (Application Programming
Interfaces) will make it easier to glue different components of the
distribution together and with external applications, as well as to
interface them to other, custom-tailored, GUIs and/or command
interpreters.

The \qe\ distribution is written, mostly, in Fortran-95, with some
parts in C or in Fortran-77. Fortran-95 offers the possibility to
introduce advanced programming techniques without sacrificing the
performances. Moreover Fortran is still the language of choice for
high-performance computing and it allows for easy integration of
legacy codes written in this language. A single source tree is used
for all architectures, with C preprocessor options selecting a small
subset of architecture-dependent code.  Parallelization is achieved
using the Message-Passing paradigm and calls to standard MPI (Message
Passing Interface) \cite{MPI} libraries. Most calls are hidden in a
few routines that act as an intermediate layer, accomplishing e.g. the
tasks of summing a distributed quantity over processors, of collecting
distributed arrays or distributing them across processors, and to
perform parallel three-dimensional Fast Fourier Transforms (FFT). This
allows to develop straightforwardly and transparently new modules and
functionalities that preserve the efficient parallelization backbone
of the codes.

\subsection{QE-forge}

The ambition of the \qe\ project is not limited to providing highly
efficient and user-friendly software for large-scale
electronic-structure calculations and materials modeling. \qe\ aims at
promoting active cooperation between a vast and diverse community of
scientists developing new methods and algorithms in
electronic-structure theory and of end users interested in their
application to the numerical simulation of materials and devices.

As mentioned, the main source of inspiration for the model we want to
promote is the successful cooperative experience of the {\em GNU/Linux
} developers' and users' community. One of the main outcomes of this
community has been the incorporation within the {\em GNU/Linux }
operating system distributions of third-party software components,
which, while being developed and maintained by autonomous, and often
very small, groups of users, are put at the disposal of the entire
community under the terms of the GPL. The community, in turn, provides
positive feedback and extensive validation by benchmarking new
developments, reporting bugs, and requesting new features.  These
developments have largely benefited from the {\em SourceForge} code
repository and software development service \cite{SF}, or by other
similar services, such as RubyForge, Tigris.org, BountySource,
BerliOS, JavaForge, and GNU Savannah.

Inspired by this model, the \qe\ developers' and users' community has
set up its own web portal, named \texttt{qe-forge}
\cite{qe-forge}. The goal of \texttt{qe-forge} is to complement the
traditional web sites of individual scientific software projects,
which are passive instruments of information retrieval, with a
dynamical space for active content creation and sharing.  Its aim is
to foster and simplify the coordination and integration of the
programming efforts of heterogeneous groups and to ease the
dissemination of the software tools thus obtained.

\texttt{qe-forge} provides, through a user-friendly web interface, an
integrated development environment, whereby researchers can freely
upload, manage and maintain their own software, while retaining full
control over it, including the right of not releasing it.  The
services so far available include source-code management software (CVS
or SVN repository), mailing lists, public forums, bug tracking
facilities, up/down-load space, and wiki pages for projects'
documentation.  \texttt{qe-forge} is expected to be the main tool by
which \qe\ end users and external contributors can maintain
\qe-related projects and make them available to the community.

\subsection{Educational usage of \qe}

Training on advanced simulation techniques using the \qe\ distribution
is regularly offered at SISSA to first-year graduate students within
the {\em electronic structure} course. The scope of this course is not
limited to the opportunities that modern simulation techniques based
on electronic-structure theory offer to molecular and materials
modeling. Emphasis is put onto the skills that are necessary to turn
new ideas into new algorithms and onto the methods that are needed to
validate the implementation and application of computer simulation
methods. Based on this experience, the \qe\ developers' group offers
on a regular basis training courses to graduate students and young
researchers worldwide, also in collaboration with the {\em Abdus
  Salam} International Centre for Theoretical Physics, which operates
under the aegis of the UNESCO and IAEA agencies of the
UNO.

% The \qe\ distribution is used in undergraduate and graduate courses in
% universities both as a tool to teach basic chemical and physical
% concepts and to provide students with an understanding of strategies,
% methods and capabilities of computer simulations in materials
% modeling.

\begin{figure}[ht]
  \hfill\hbox{\includegraphics[width=0.8\columnwidth]{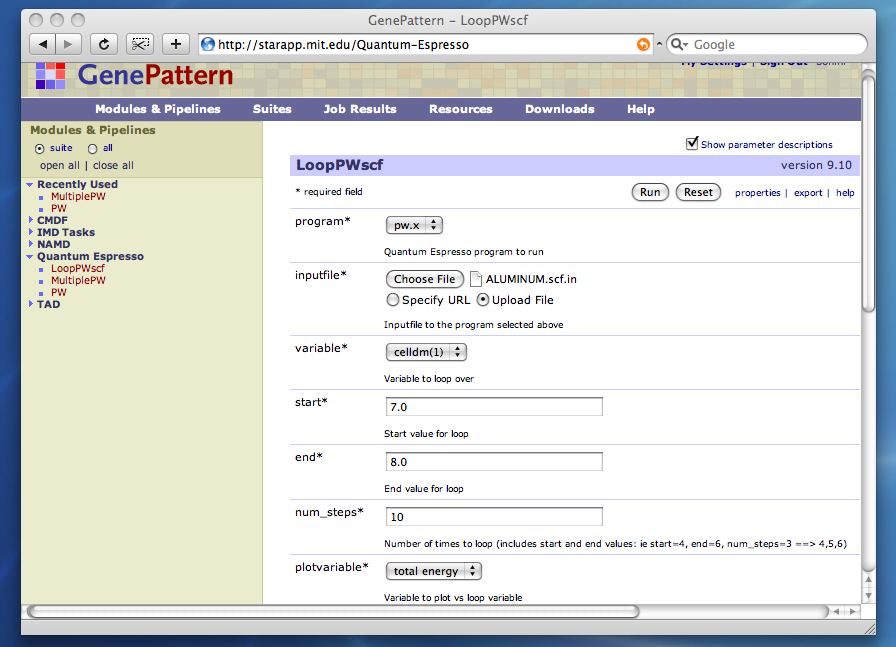}\quad}
  \caption{Snapshot of the web interface used for undergraduate
    teaching at MIT. The software has been developed at the MIT's
    Office of Educational Innovation and Technology.}
  \label{fig:webinterface}
\end{figure}

The \qe\ distribution is used not only for graduate, but also for
undergraduate training. At MIT, for example, it is one of the teaching
tools in the class {\em Introduction to Modeling and Simulations}---an
institute-wide course offered to undergraduates from the School of
Science and the School of Engineering. The challenge here is to
provide students of different backgrounds with an overview of
numerical simulations methods to study properties of real materials.
For many undergraduates, this represents the first experience of
computers used as scientific tools.  To facilitate the access and use
of \qe\, a user-friendly web interface has been developed at MIT,
based on the GenePattern portal, that allows direct access to the
code, thus removing the need to use a Unix/Linux environment or the
details of the job queueing and submission procedure. The user
utilizes a web browser (see Fig. \ref{fig:webinterface}) to build
input files and view the outputs of simulations, and to perform
calculations from wherever Internet access is available. The
calculations run on dedicated computer clusters where the code has
been previously installed and tested.

Using a web interface to easily access computational resources and
share them among different users naturally points to the concept of
cloud computing, and the previous model was tested at MIT in the
Spring 2009, wholly based on a cluster of virtual machines on Amazon's
Elastic Compute Cloud (EC2) web service. Our experience shows that
when compared to the cost of purchasing, maintaining and administering
computer clusters, the use of web-based computational resources
becomes a very appealing and affordable option. It is particularly
suited for classroom instruction, where advanced computational
performance is not required, and it allows for easy transferability of
this resource across universities.

\section{Short description of \qe}

\qe\ implements a variety of methods and algorithms aimed at a
chemically realistic modeling of materials from the nanoscale upwards,
based on the solution of the density-functional theory (DFT)
\cite{dft1,dft2} problem, using a plane waves (PW) basis set and
pseudopotentials (PP) \cite{pwpp} to represent electron-ion
interactions.

The codes are constructed around the use of periodic boundary
conditions, which allows for a straightforward treatment of infinite
crystalline systems, and an efficient convergence to the thermodynamic
limit for aperiodic but extended systems, such as liquids or amorphous
materials.  Finite systems are also treated using supercells; if
required, open-boundary conditions can be used through the use of the
density-countercharge method \cite{ismaila08}. \qe\ can thus be used
for any crystal structure or supercell, and for metals as well as for
insulators.  The atomic cores can be described by separable \cite{KB}
norm-conserving (NC) PPs \cite{HSC}, ultra-soft (US) PPs \cite{USPP},
or by projector-augmented wave (PAW) sets \cite{PAW1}.  Many different
exchange-correlation functionals are available in the framework of the
local-density (LDA) or generalized-gradient approximation (GGA)
\cite{GGA}, plus advanced functionals like Hubbard $U$ corrections and
few meta-GGA \cite{TPSS} and hybrid functionals
\cite{PBE0,Becke1,B3LYP}.  The latter is an area of very active
development, and more details on the implementation of hybrid
functionals and related Fock-exchange techniques are given in Appendix
A.5.
 
The basic computations/simulations that can be performed include:
\begin{itemize}
\item calculation of the Kohn-Sham (KS) orbitals and energies \cite{KS} for
  isolated or extended/periodic systems, and of their ground-state
  energies;
\item complete structural optimizations of the microscopic (atomic
  coordinates) and macroscopic (unit cell) degrees of freedom, using
  Hellmann-Feynman forces \cite{Hellmann,Feynman} and stresses
  \cite{stress};
\item ground states for magnetic or spin-polarized system, including
  spin-orbit coupling \cite{relativistic} and non-collinear magnetism
  \cite{noncol1,noncol2};
\item {\em ab initio} molecular dynamics (MD), using either the
  Car-Parrinello Lagrangian \cite{CP} or the Hellmann-Feynman forces
  calculated on the Born-Oppenheimer (BO) surface \cite{BOMD}, in a
  variety of thermodynamical ensembles, including NPT variable-cell
  \cite{vcmd0,vcmd2} MD;
\item density-functional perturbation theory (DFPT)
  \cite{bgt,dfpt1,dfpt2}, to calculate second and third derivatives of
  the total energy at any arbitrary wavelength, providing phonon
  dispersions, electron-phonon and phonon-phonon interactions, and
  static response functions (dielectric tensors, Born effective
  charges, IR spectra, Raman tensors);
\item location of saddle points and transition states via
  transition-path optimization using the nudged elastic band (NEB)
  method \cite{neb1,neb2,neb3};
\item ballistic conductance within the Landauer-B\"uttiker theory
  using the scattering approach \cite{ball1};
\item generation of maximally localized Wannier functions
  \cite{MaxWan1,MaxWan2} and related quantities;
\item calculation of nuclear magnetic resonance (NMR) and electronic
  paramagnetic resonance (EPR) parameters
  \cite{Pickard_2001_a_gipaw_PRB,Pickard_2003_a_gipaw_PRL};
\item calculation of K-edge X-ray absorption spectra
  \cite{Taillefumier}.
\end{itemize}
Other more advanced or specialized capabilities are described in the
next sections, while ongoing projects (e.g. time-dependent DFT and
many-body perturbation theory) are mentioned in the last section.
Selected applications were described in Ref~\cite{ZK}.  Several
utilities for data post-processing and interfacing to advanced graphic
applications are available, allowing e.g. to calculate scanning
tunneling microscopy (STM) images \cite{stm}, the electron
localization function (ELF) \cite{elf}, L\"owdin charges \cite{QC},
the density of states (DOS), and planar \cite{macroscopic} or
spherical averages of the charge and spin densities and potentials.

\subsection{Data file format}

The interoperability of different software components within a complex
project such as \qe\ relies on the careful design of file formats for
data exchange. A rational and open approach to data file formats is
also essential for interfacing applications within \qe\ with
third-party applications, and more generally to make the results of
lengthy and expensive computer simulations accessible to, and
reproducible by, the scientific community at large. The need for data
file formats that make data exchange easier than it is now is starting
to be widely appreciated in the electronic-structure community.  This
problem has many aspects and likely no simple, "one-size-fits-all",
solution. Data files should ideally be
\begin{itemize}
\item {\em extensible}: one should be able to add some more
  information to a file without breaking all codes that read that
  file;
\item {\em self-documenting}: it should be possible to understand the
  contents of a file without too much effort;
\item {\em efficient}: with data size in the order of GBytes for
  large-scale calculations, slow or wasteful I/O should be avoided.
\end{itemize}
The current trend in the electronic-structure community seems to be
the adoption of one of the following approaches:
\begin{itemize}
\item Structured file formats, notably Hierarchical Data Format (HDF)
  \cite{hdf} and network Common Data Form (netCDF) \cite{netcdf}, that
  have been widely used for years in other communities;
\item file formats based on the Extensible Markup Language (XML)
  \cite{xml}.
\end{itemize}
It is unlikely that a common, standardized data format will ever
prevail in our community. We feel that we should focus, rather than on
standardization, on an approach that allows an easy design and usage
of simple and reliable converters among different data
formats. Prompted by these considerations, \qe\ developers have opted
for a simple solution that tries to combine the advantages of both the
above-mentioned approaches.  A single file containing all the data of
a simulation is replaced by a {\em data directory}, containing several
files and subdirectories, much in the same way as it is done in the
Mac OS X operating system. The ``head'' file contains data written
with ordinary Fortran formatted I/O, identified by XML tags.  Only
data of small size, such as atomic positions, parameters used in the
calculation, one-electron and total energies, are written in the head
file. Data of potentially large size, such as PW coefficients of
KS orbitals, charge density, and potentials, are present as links
to separate files, written using unformatted Fortran I/O. Data for
each {\bf k}-point are written to a separate subdirectory.  A
lightweight library called \texttt{iotk}, standing for Input/Output
ToolKit~\cite{iotk}, is used to read and write the data directory.

Another problem affecting interoperability of PW-PP codes is the
availability of data files containing atomic PP's---one of the basic
ingredients of the calculation.  There are many different types of
PP's, many different codes generating PP's (see e.g. Ref
\cite{fhi98PP,VdB-uspp,opium}), each one with its own format. Again,
the choice has fallen on a simple solution that makes it easy to write
converters from and to the format used by \qe. Each atomic PP is
contained in a formatted file (efficiency is not an issue here),
described by a XML-like syntax. The resulting format has been named
Unified Pseudopotential File (UPF). Several converters from other
formats to the UPF format are available in \qe.

\section{\qe\ packages}

The complete \qe\ distribution is rather large. The current 4.1 version
includes about 310,000 lines of Fortran-90 code, 1,000 lines of Fortran-77
code, 1,000 lines of C code, 2000 lines of Tcl code, plus parts of external
libraries such as FFTW, BLAS, LAPACK and the external toolkit \texttt{iotk}. 
In addition, there are approx. 10,000 lines of specific documentation (not 
counting different formats), more than 100 different examples and more than
100 tests of the different functionalities. Overall the complete distribution
includes more than 3000 files, organized into 200 directories, and takes
22Mb in compressed format.

With such a sizable code basis, modularization becomes necessary. \qe\ is
presently divided into several executables, performing different types
of calculations, although some of them have overlapping
functionalities. Typically there is a single set of
functions/subroutines or a single Fortran 90 module that performs each
specific task (e.g. matrix diagonalizations, or potential updates),
but there are still important exceptions to this rule, reflecting the
different origin and different styles of the original components. \qe\
has in fact been built out of the merge and re-engineering of
different packages, that were independently developed for several
years. In the following, the main components are briefly described.

\subsection{PWscf}

\label{sec:PWscf}
\texttt{PWscf} implements an iterative approach to reach
self-consistency, using at each step iterative diagonalization
techniques, in the framework of the plane-wave pseudopotential
method. An early version of \texttt{PWscf} is described in
Ref~\cite{PWscf}.

Both separable NC-PPs and US-PPs are implemented; recently, also the
projector augmented-wave method \cite{PAW1} has been added, largely
following the lines of Ref~\cite{PAW2} for its implementation.  In the
case of US-PPs, the electronic wave functions can be made smoother at
the price of having to augment their square modulus with additional
contributions to recover the actual physical charge densities. For
this reason, the charge density is more structured than the square of
the wavefunctions, and requires a larger energy cutoff for its plane
wave expansion (typically, 6 to 12 times larger; for a NC-PP, a factor
of 4 would be mathematically sufficient).  Hence, different real-space
Fourier grids are introduced - a "soft" one that represents the square
of electronic wave functions, and a "hard" one that represents the
charge density \cite{carpasqua1,carpasqua2}.  The augmentation terms
can be added either in reciprocal space (using an exact but expensive
algorithm) or directly in real space (using an approximate but faster
algorithm that exploits the local character of the augmentation
charges).

\texttt{PWscf} can use the well established LDA and GGA
exchange-correlation functionals, including spin-polarization within
the scheme proposed in Ref~\cite{WhiteBird} and can treat
non-collinear magnetism\cite{noncol1,noncol2} as e.g.  induced by
relativistic effects (spin-orbit interactions)
\cite{spinorbit1,spinorbit2} or by complex magnetic interactions ({\em
  e.g.} in the presence of frustration). DFT + Hubbard U calculations
\cite{LDA+U} are implemented for a simplified (``no-$J$'')
rotationally invariant form \cite{Ucococcioni} of the Hubbard
term. Other advanced functionals include TPSS meta-GGA \cite{TPSS},
functionals with finite-size corrections \cite{KZK}, and the PBE0
\cite{PBE0} and B3LYP \cite{Becke1,B3LYP} hybrids.

Self-consistency is achieved via the modified Broyden method of
Ref~\cite{scf}, with some further refinements that are detailed in
Appendix A.1.  The sampling of the Brillouin Zone (BZ) can be performed
using either special \cite{ChadiCohen,MonkhorstPack} \textbf{k}-points
provided in input or those automatically calculated starting from a
uniform grid.  Crystal symmetries are automatically detected and
exploited to reduce computational costs, by restricting the sampling
of the BZ to the irreducible wedge alone (See Appendix A.4). 
 When only the $\Gamma$
point (${\bf k}=0$) is used, advantage is taken of the real character
of the KS orbitals, allowing one to store just half of the Fourier
components. BZ integrations in metallic systems can be performed using
a variety of smearing/broadening techniques, such as Fermi-Dirac, 
Gaussian, Methfessel-Paxton \cite{MP}, and Marzari-Vanderbilt cold 
smearing \cite{cold}. The tetrahedron method \cite{tetra} is also
implemented. Finite-temperature effects on the electronic
properties can be easily accounted for by using the Fermi-Dirac
smearing as a practical way of implementing the Mermin finite-temperature
density-functional approach \cite{mermin}.

Structural optimizations are performed using the
Broyden-Fletcher-Goldfarb-Shanno (BFGS) algorithm
\cite{opt,bfgs1,bfgs2} or damped dynamics; these can involve both the
internal, microscopic degrees of freedom (i.e. the atomic coordinates)
and/or the macroscopic ones (shape and size of the unit cell).  The
calculation of minimum-energy paths, activation energies, and
transition states uses the Nudged Elastic Band (NEB) method
\cite{neb1}. Potential energy surfaces as a function of suitably
chosen collective variables can be studied using Laio-Parrinello
metadynamics \cite{metadyn}.

Microcanonical (NVE) MD is performed on the BO surface, {\em i.e.}
achieving electron self-consistency at each time step, using the
Verlet algorithm\cite{Verlet}.  Canonical (NVT) dynamics can be
performed using velocity rescaling, or Anderson's or Berendsen's
thermostats \cite{allentild}. Constant-pressure (NPT) MD is performed
by adding additional degrees of freedom for the cell size and volume,
using either the Parrinello-Rahman Lagrangian \cite{vcmd1} or the
so-called {\em
  invariant} Lagrangian of Wentzcovitch \cite{vcmd2}.

The effects of finite macroscopic electric fields on the electronic
structure of the ground state can be accounted for either through the
method of Ref~\cite{Efield,souza-vanderbilt} based on the Berry phase,
or (for slab geometries only) through a sawtooth external potential
\cite{kunc,tobik}.  A quantum fragment can be embedded in a complex
electrostatic environment that includes a model solvent
\cite{damian06} and a counterion distribution \cite{ismailaec09}, as
is typical of electrochemical systems.

\subsection{CP}

\label{sec:CP}
The \texttt{CP} code is the specialized module performing
Car-Parrinello {\em ab initio} MD.  \texttt{CP} can use both NC PPs
\cite{carpar1} and US PPs \cite{carpasqua0,carpasqua1}. In the latter
case, the electron density is augmented through a Fourier
interpolation scheme in real space (``box grid'')
\cite{carpasqua1,carpasqua2} that is particularly efficient for large
scale calculations.  \texttt{CP} implements the same functionals as
\texttt{PWscf}, with the exception of hybrid functionals; a simplified
one-electron self-interaction correction (SIC)\cite{sic} is also
available.  The Car-Parrinello Lagrangian can be augmented with
Hubbard U corrections \cite{sit07}, or Hubbard-based penalty
functionals to impose arbitrary oxidation states \cite{sit06}.

Since the main applications of \texttt{CP} are for large systems
without translational symmetry (e.g. liquids, amorphous materials),
Brillouin zone sampling is restricted to the $\Gamma$ point of the
supercell, allowing for real instead of complex
wavefunctions. Metallic systems can be treated in the framework of
``ensemble DFT'' \cite{ensemble}.

In the Car-Parrinello algorithm, microcanonical (NVE) MD is performed
on both electronic and nuclear degrees of freedom, treated on the same
footing, using the Verlet algorithm. The electronic equations of
motion are accelerated through a preconditioning scheme \cite{cppre}.
Constant-pressure (NPT) MD is performed using the Parrinello-Rahman
Lagrangian \cite{vcmd1} and additional degrees of freedom for the
cell.  Nos\'e-Hoover thermostats \cite{Nose} and Nos\'e-Hoover chains
\cite{martyna:2635} allow to perform simulations in the different
canonical ensembles.

\texttt{CP} can also be used to directly minimize the energy
functional to self-consistency while keeping the nuclei fixed, or to
perform structural minimizations of nuclear positions, using the
``global minimization'' approaches of 
Refs. \cite{payne,marxhutter}, and damped dynamics or
conjugate-gradients on the electronic or ionic degrees of freedom.  It
can also perform NEB and metadynamics calculations.

Finite homogeneous electric fields can be accounted for using the
Berry phase method, adapted to systems with the $\Gamma$ point only
\cite{Efield}. This advanced feature can be used in combination with
MD to obtain the infrared spectra of liquids \cite{Efield,Dubois}, the
low- and high-frequency dielectric constants \cite{Efield,Giustino}
and the coupling factors required for the calculation of vibrational
properties, including infrared, Raman
\cite{IRandRaman,IRandRaman2,IRandRaman3}, and hyper-Raman
\cite{HyperRaman} spectra.

\subsection{PHonon}

The \texttt{PHonon} package implements density-functional perturbation
theory (DFPT) \cite{bgt,dfpt1,dfpt2} for the calculation of second-
and third-order derivatives of the energy with respect to atomic
displacements and to electric fields.  The global minimization
approach \cite{c60ph,dfpt3} is used for the special case of normal
modes in finite (molecular) systems, where no BZ sampling is required
(\texttt{Gamma} code). In the general case a self-consistent procedure
\cite{dfpt1} is used, with the distinctive advantage that the response
to a perturbation of any arbitrary wavelength can be calculated with a
computational cost that is proportional to that of the unperturbed
system. Thus, the response at any wavevector, including very small
(long-wavelength) ones, can be inexpensively calculated.  This latter
approach, and the technicalities involved in the calculation of
effective charges and interatomic force constants, are described in
detail in Refs. \cite{dfpt1,dfpt4} and implemented in the \texttt{PH} code.

Symmetry is fully exploited in order to reduce the amount of
computation. Lattice distortions transforming according to irreducible
representations of small dimensions are generated first. The
charge-density response to these lattice distortions is then sampled
at a number of discrete \textbf{k}-points in the BZ, which is reduced
according to the symmetry of the small group of the phonon wavevector
\textbf{q}. The grid of the \textbf{q} points needed for the
calculation of interatomic force constants reduces to one wavevector
per star: the dynamical matrices at the other \textbf{q} vectors in
the star are generated using the symmetry operations of the crystal.
This approach allows us to speed up the calculation without the need
to store too much data for symmetrization.
 
The calculation of second-order derivatives of the energy works also
for US PP \cite{Dalcorso1,Dalcorso1b} and for all GGA flavors
\cite{dfptgga1,dfptgga2} used in \texttt{PWscf} and in \texttt{CP}.
The extension of \texttt{PHonon} to PAW \cite{Dalcorso_paw}, to
noncollinear magnetism and to fully relativistic US PPs which include
spin-orbit coupling \cite{Dalcorso_so} will be available by the time
this paper will be printed.

Advanced features of the \texttt{PHonon} package include the calculation
of third order derivatives of the energy, such as electron-phonon or
phonon-phonon interaction coefficients.  Electron-phonon interactions
are straightforwardly calculated from the response of the
self-consistent potential to a lattice distortion. This involves a
numerically-sensitive ``double-delta'' integration at the Fermi
energy, that is performed using interpolations on a dense
\textbf{k}-point grid. Interpolation techniques based on Wannier
functions \cite{giustino07} will speed up considerably these
calculations.  The calculation of the anharmonic force constants from
third-order derivatives of the electronic ground-state energy is
described in Ref.~\cite{anharm} and is performed by a separate code
called \texttt{d3}. Static Raman coefficients are calculated using the
second-order response approach of Refs. \cite{raman1,raman2}. Both
third-order derivatives and Raman-coefficients calculations are
currently implemented only for NC PPs.

\subsection{atomic}

The \texttt{atomic} code performs three different tasks: {\em i)}
solution of the self-consistent all-electron radial KS equations (with
a Coulomb nuclear potential and spherically symmetric charge density);
{\em ii)} generation of NC PPs, of US PPs, or of PAW data-sets; {\em
  iii)} test of the above PPs and data-sets. These three tasks can be
either separately executed or performed in a single run.  Three
different all-electron equations are available: {\em i)} the non
relativistic radial KS equations, {\em ii)} the scalar relativistic
approximation to the radial Dirac equations~\cite{KH}, {\em iii)} the
radial Dirac-like equations derived within relativistic density
functional theory~\cite{RDFT,RDFT2}.  For {\em i)} and {\em ii)}
atomic magnetism is dealt with within the local spin density
approximation, i.e. assuming an axis of magnetization. The
\texttt{atomic} code uses the same exchange and correlation energy
routines of \texttt{PWscf} and can deal with the same functionals.

The code is able to generate NC PPs directly in separable form (also
with multiple projectors per angular momentum channel) via the
Troullier-Martins~\cite{TM} or the
Rappe-Rabe-Kaxiras-Joannopoulos~\cite{RRKJ} pseudization.  US PPs can
be generated by a two-step pseudization process, starting from a NC
PPs, as described in Ref.~\cite{Kresse_pseudi}, or using the solutions
of the all-electron equation and pseudizing the augmentation
functions~\cite{carpasqua1}.  The latter method is used also for the
PAW data-set generation.  The generation of fully relativistic NC and
US PPs including spin-orbit coupling effects is also available.
Converters are available to translate pseudopotentials encoded in
different formats (e.g. according to the Fritz-Haber \cite{fhi98PP} or
Vanderbilt \cite{VdB-uspp} conventions) into the UPF format adopted by
\qe.

Transferability tests can be made simultaneously for several atomic
configurations with or without spin-polarization, by solving the non
relativistic radial KS equations generalized for separable
nonlocal PPs and for the presence of an overlap matrix.

\subsection{PWcond}

The \texttt{PWcond} code implements the scattering approach proposed
by Choi and Ihm \cite{ball1} for the study of coherent electron
transport in atomic-sized nanocontacts within the Landauer-B\"uttiker
theory. Within this scheme the linear response ballistic conductance
is proportional to the quantum-mechanical electron transmission at the
Fermi energy for an open quantum system consisting of a scattering
region (e.g., an atomic chain or a molecule with some portions of left
and right leads) connected ideally from both sides to semi-infinite
metallic leads. The transmission is evaluated by solving the KS
equations with the boundary conditions that an electron coming from
the left lead and propagating rightwards gets partially reflected and
partially transmitted by the scattering region. The total transmission
is obtained by summing all transmission probabilities for all the
propagating channels in the left lead. As a byproduct of the method,
the \texttt{PWcond} code provides the complex band structures of the
leads, that is the Bloch states with complex $k_z$ in the direction of
transport, describing wave functions exponentially growing or decaying
in the $z$ direction.  The original method formulated with NC PPs has
been generalized to US PPs both in the scalar relativistic
\cite{ball2} and in the fully relativistic forms \cite{ball3}.

\subsection{GIPAW}

The \texttt{GIPAW} code allows for the calculation of physical
parameters measured by {\em i)} nuclear magnetic resonance (NMR) in
insulators (the electric field gradient (EFG) tensors and the chemical
shift tensors), and by {\em ii)} electronic paramagnetic resonance
(EPR) spectroscopy for paramagnetic defects in solids or in radicals
(the hyperfine tensors and the g-tensor). The code also computes the
magnetic susceptibility of nonmagnetic insulators. \texttt{GIPAW} is based
on the PW-PP method, and uses many subroutines of \texttt{PWscf} and
of \texttt{PHonon}. The code is currently restricted to NC PPs. All
the NMR and EPR parameters depend on the detailed shape of the
electronic wave-functions near the nuclei and thus require the
reconstruction of the all-electron wave-functions from the PP
wave-functions. For the properties defined at zero external magnetic
field, namely the EFG and the hyperfine tensors, such reconstruction
is performed as a post-processing step of a self-consistent
calculation using the PAW reconstruction, as described for the EFG in
Ref. \cite{Profeta_2003_a} and for the hyperfine tensor in
Ref. \cite{vandeWalle_1993_a}. The g-tensor, the NMR chemical shifts
and the magnetic susceptibility are obtained from the orbital linear
response to an external uniform magnetic field. In the presence of a
magnetic field the PAW method is no more gauge- and translationally
invariant. Gauge and translational invariances are restored by using
the gauge including projector augmented wave (GIPAW) method
\cite{Pickard_2001_a_gipaw_PRB,Pickard_2003_a_gipaw_PRL} both {\em i)}
to describe in the PP Hamiltonian the coupling of orbital degrees of
freedom with the external magnetic field, and {\em ii)} to reconstruct
the all-electron wave-functions, in presence of the external magnetic
field. In addition, the description of a uniform magnetic field within
periodic boundary conditions is achieved by considering the long
wave-length limit of a sinusoidally modulated field in real space
\cite{Mauri_1996_a,Mauri_1996_b}. The NMR chemical shifts are computed
following the method described in Ref. \cite{Pickard_2001_a_gipaw_PRB},
the g-tensor following Ref. \cite{Pickard_2002_a_gtensor} and the
magnetic susceptibility following
Refs. \cite{Mauri_1996_a,Pickard_2001_a_gipaw_PRB}.  Recently, a
``converse'' approach to calculate chemical shifts has also been
introduced \cite{converseNMR}, based on recent developments on the
Berry-phase theory of orbital magnetization; since it does not require
a linear-response calculation, it can be straightforwardly applied to
arbitrarily complex exchange-correlation functionals, and to very
large systems, albeit at a computational cost that is proportional to
the number of chemical shifts that need to be calculated.

\subsection{XSPECTRA}

The \texttt{XSPECTRA} code allows for the calculation of K-edge X-ray
absorption spectra (XAS).  The code calculates the XAS cross-section 
including both dipolar and quadrupolar matrix elements.  The
code uses the self-consistent charge density produced by
\texttt{PWscf} and acts as a post-processing tool.  The all-electron
wavefunction is reconstructed using the PAW method and its
implementation in the \texttt{GIPAW} code.  The presence of a
core-hole in the final state of the X-ray absorption process is
simulated by using a pseudopotential for the absorbing atom with a
hole in the 1s state. The calculation of the charge density is
performed on a supercell with one absorbing atom.  From the
self-consistent charge density, the X-ray absorption spectra are
obtained using the Lanczos method and a continued fraction expansion
\cite{Taillefumier,GougoussisUS}. The advantage of this approach is
that once the charge density is known it is not necessary to calculate
empty bands to describe very high energy features of the spectrum.
Correlation effects can be simulated in a mean-field way using the
Hubbard U correction \cite{Ucococcioni} that has been included in the
\texttt{XSPECTRA} code in Ref. \cite{Gougoussis}.  Currently the code
is limited to collinear magnetism.  Its extension to noncollinear
magnetism is under development.

\subsection{Wannier90}

\texttt{Wannier90} \cite{wannier90, wannier90-url} is a code that
calculates maximally-localized Wannier functions in insulators or
metals---according to the algorithms described in
Refs. \cite{MaxWan1,MaxWan2}---and a number of properties that can be
conveniently expressed in a Wannier basis. The code is developed and
maintained independently by a Wannier development group
\cite{wannier90-url} and can be taken as a representative example of
the philosophy described earlier, where a project maintains its own
individual distribution but provides full interoperability with the
core components of \qe\, in this case \texttt{PWscf} or
\texttt{CP}. These codes are in fact used as ``quantum engines'' to
produce the data onto which \texttt{Wannier90} operates.  The need to
provide transparent protocols for interoperability has in turn
facilitated the interfacing of \texttt{wannier90} with other quantum
engines \cite{abinit,castep}, fostering a collaborative engagement
with the broader electronic-structure community that is also in the
spirit of \qe\ .

\texttt{Wannier90} requires as input the scalar products between
wavefunctions at neighboring k-points, where these latter form uniform
meshes in the Brillouin zone. Often, it is also convenient to provide
scalar products between wavefunctions and trial, localized real-space
orbitals---these are used to guide the localization procedure towards
a desired, physical minimum. As such, the code is not tied to a
representation of the wavefunctions in any particular basis---for
\texttt{PWscf} and \texttt{CP} a post-processing utility is in charge
of calculating these scalar products using the plane-wave basis set of
\qe\ and either NC-PPs or US-PPs. Whenever $\Gamma$ sampling is used,
the simplified algorithm of Ref. \cite{pierino98} is adopted.

Besides calculating maximally localized Wannier functions, the code is
able to construct the Hamiltonian matrix in this localized basis,
providing a chemically accurate, and transferable, tight-binding
representation of the electronic structure of the system. This, in
turn, can be used to construct Green's functions and self-energies for
ballistic transport calculations \cite{calzolari04,yslee05}, to
determine the electronic structure and DOS of very large
scale structures \cite{yslee05}, to interpolate accurately the
electronic band structure (i.e. the Hamiltonian) across the Brillouin
zone \cite{yslee05,yates07}, or to interpolate any other operator
\cite{yates07}.  These latter capabilities are especially useful for
the calculation of integrals that depend sensitively on a submanifold
of states; common examples come from properties that depend
sensitively on the Fermi surface, such as electronic conductivity,
electron-phonon couplings Knight shifts, or the anomalous Hall
effect. A related by-product of \texttt{Wannier90} is the capability
of downfolding a selected, physically significant manifold of bands
into a minimal but accurate basis, to be used for model Hamiltonians
that can be treated with complex many-body approaches.

\subsection{PostProc}

The \texttt{PostProc} module contains a number of codes for
post-processing and analysis of data files produced by \texttt{PWscf}
and \texttt{CP}.  The following operations can be performed:
\begin{itemize}
\item Interfacing to graphical and molecular graphics
  applications. Charge and spin density, potentials, ELF \cite{elf}
  and STM images \cite{stm} are extracted or calculated and written to
  files that can be directly read by most common plotting programs,
  like \texttt{xcrysden} \cite{xcrysden} and \texttt{VMD}
  \cite{Hump96}.
\item Interfaces to other codes that use DFT results from \qe\ for
  further calculations, such as e.g.: \texttt{pw2wannier90}, an
  interface to the \texttt{wannier90} library and code
  \cite{wannier90,wannier90-url} (also included in the \qe\
  distribution); \texttt{pw2casino.f90}, an interface to the
  \texttt{casino} quantum Monte Carlo code \cite{casino};
  \texttt{wannier\_ham.f90}, a tool to build a tight-binding
  representation of the KS Hamiltonian to be used by the \texttt{dmft}
  code \cite{dmft} (available at the \texttt{qe-forge} site);
  \texttt{pw\_export.f90}, an interface to the GW code \texttt{SaX}
  \cite{sax}; \texttt{pw2gw.f90}, an interface to code \texttt{DP}
  \cite{dp} for dielectric property calculations, and to code
  \texttt{EXC} \cite{exc} for excited-state properties.
\item Calculation of various quantities that are useful for the
  analysis of the results. In addition to the already mentioned ELF
  and STM, one can calculate projections over atomic states
  (e.g. L\"owdin charges \cite{QC}), DOS and Projected DOS (PDOS),
  planar and spherical averages, and the complex macroscopic
  dielectric function in the random-phase approximation (RPA).
\end{itemize}

\def\pwgui{\texttt{PWgui}}

\begin{figure*}[ht]
  \centering
  \setbox0=\hbox{
    \includegraphics[width=0.47\columnwidth]{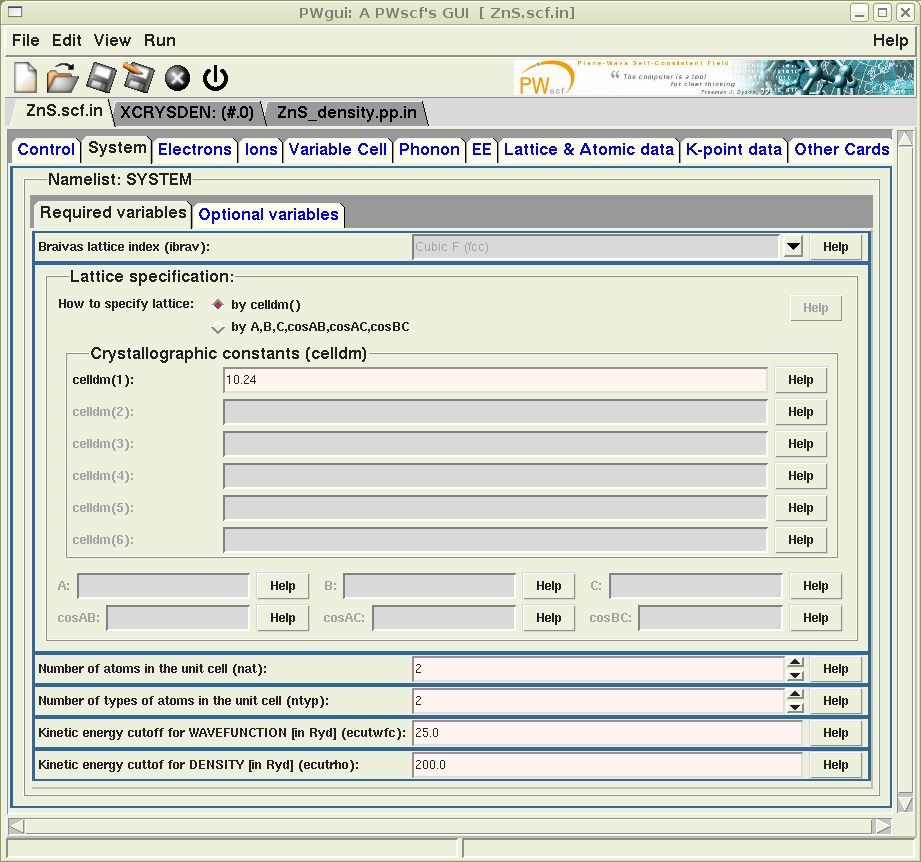}
  }
  \setbox1=\hbox{
    \includegraphics[width=0.32\columnwidth]{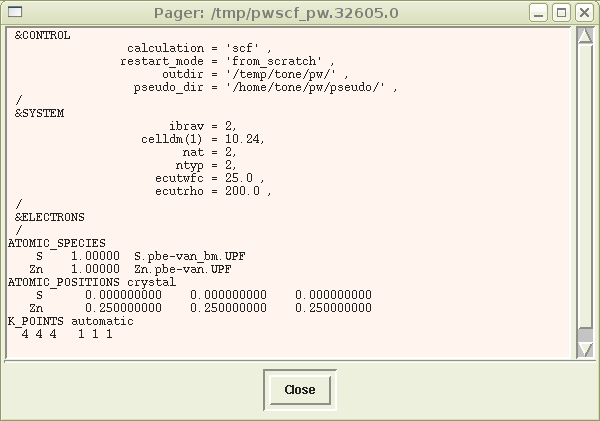}
  }
  \hfill\hbox to 0.8\columnwidth{\hfill\copy0\hfill
    \vbox to \ht0{
      \vfill \box1 \vfill
    }
    \quad
  }
  
  \caption{Snapshot of the \pwgui\ application. Left: \pwgui's main
    window; right: preview of specified input data in text mode.}
  \label{fig:pwgui}
\end{figure*}

\subsection{PWgui}

\pwgui\ is the graphical user interface (GUI) for the \texttt{PWscf},
\texttt{PHonon}, and \texttt{atomic} packages as well as for some of
the main codes in \texttt{PostProc} (e.g. \texttt{pp.x} and
\texttt{projwfc.x}).  \pwgui\ is an input file builder whose main goal
is to lower the learning barrier for the newcomer, who has to struggle
with the input syntax. Its event-driven mechanism automatically
adjusts the display of required input fields (i.e. enables certain
sets of widgets and disables others) to the specific cases selected
(see Fig. \ref{fig:pwgui}, left panel). It enables a preview of the
format of the (required) input file records for a given type of
calculation (see Fig. \ref{fig:pwgui}, right panel).  The input files
created by \pwgui\ are guaranteed to be syntactically correct
(although they can still be physically meaningless). It is possible to
upload previously generated input files for syntax checking and/or to
modify them.  It is also possible to run calculations from within the
\pwgui. In addition, \pwgui\ can also use the external
\texttt{xcrysden} program \cite{xcrysden} for the visualization of
molecular and/or crystal structures from the specified input data and
for the visualization of properties (e.g. charge densities or STM
images).

As the \qe\ codes evolve, the input file syntax expands as well. This
implies that \pwgui\ has to be continuously adapted. To effectively
deal with such issue, \pwgui\ uses the \texttt{GUIB} concept
\cite{guib}. \texttt{GUIB} builds on the consideration that the input
files for numerical simulation codes have a rather simple structure
and it exploits this simplicity by defining a special meta-language
with two purposes: the first is to define the input-file syntax, and
the second is to simultaneously automate the construction of the GUI
on the basis of such a definition.

A similar strategy has been recently adopted for the description of
the \qe\ input file formats. A single definition/description of a
given input file serves {\em i)} as a documentation {\em per-se}, {\em
  ii)} as a \pwgui\ help documentation, and {\em iii)} as a utility to
synchronize the \pwgui\ with up-to-date input file formats.

\section{Parallelization}

Keeping the pace with the evolution of high-end supercomputers is one
of the guiding lines in the design of \qe, with a significant effort
being dedicated to porting it to the latest available
architectures. This effort is motivated not only by the need to stay
at the forefront of architectural innovation for large to very-large
scale materials science simulations, but also by the speed at which
hardware features specifically designed for supercomputers find their
way into commodity computers.

\begin{table*}
  \caption{Summary of parallelization levels in \qe.}
  \label{tab1}
  \begin{tabular}{llll}
    group        & distributed quantities  & communications & performance \\
    \hline
    {\em image}  & NEB images  & very low       & linear CPU scaling,\\
    &             &                & fair to good load balancing;\\
    &             &                & does not distribute RAM\\
    {\em pool}   & ${\bf k}$-points & low       & almost linear CPU scaling,\\
    &             &                & fair to good load balancing;\\
    &             &                & does not distribute RAM\\
    {\em plane-wave}  & plane waves, {\bf G}-vector &high& good CPU scaling,\\
    & coefficients, {\bf R}-space  &                & good load balancing,\\
    & FFT arrays      &                & distributes most RAM\\
    \hline
    {\em task}   & FFT on electron states & high   & improves load balancing\\
    {\em linear algebra}& subspace Hamiltonians & very high & improves scaling,\\
    & and constraints matrices &       & distributes more RAM\\
    \hline
  \end{tabular}
\end{table*}

The architecture of today's supercomputers is characterized by
multiple levels and layers of inter-processor communication: the
bottom layer is the one affecting the instruction set of a single core
(simultaneous multithreading, hyperthreading); then one has parallel
processing at processor level (many CPU cores inside a single
processor sharing caches) and at node level (many processors sharing
the same memory inside the node); at the top level, many nodes are
finally interconnected with a high-performance network. The main
components of the \qe\ distribution are designed to exploit this
highly structured hardware hierarchy. High performance on massively
parallel architectures is achieved by distributing both data and
computations in a hierarchical way across available processors, ending
up with multiple parallelization levels \cite{rimini08} that can be
tuned to the specific application and to the specific
architecture. This remarkable characteristic makes it possible for the
main codes of the distribution to run in parallel on most or all
parallel machines with very good performance in all cases.

More in detail, the various parallelization levels are geared into a
hierarchy of processor groups, identified by different MPI
communicators. In this hierarchy, groups implementing coarser-grained
parallel tasks are split into groups implementing finer-grained
parallel tasks. The first level is {\em image} parallelization,
implemented by dividing processors into $n_{image}$ groups, each
taking care of one or more images (i.e. a point in the configuration
space, used by the NEB method).  The second level is {\em pool}
parallelization, implemented by further dividing each group of
processors into $n_{pool}$ pools of processors, each taking care of
one or more {\bf k}-points. The third level is {\em
  plane-wave} parallelization, implemented by distributing real- and
reciprocal-space grids across the $n_{PW}$ processors of each pool.
The final level is {\em task group} parallelization \cite{tg}, in
which processors are divided into $n_{task}$ task groups of
$n_{FFT}=n_{PW}/n_{task}$ processors, each one taking care of
different groups of electron states to be Fourier-transformed, while
each FFT is parallelized inside a task group. A further
paralellization level, {\em linear-algebra}, coexists side-to-side
with plane-wave parallelization, {\em i.e.} they take care of
different sets of operations, with different data
distribution. Linear-algebra parallelization is implemented both with
custom algorithms and using ScaLAPACK \cite{scalapack}, which on
massively parallel machines yield much superior performances.  Table
\ref{tab1} contains a summary of the five levels currently
implemented.  With the recent addition of the two last levels, most
parallelization bottlenecks have been removed, while both computations
and data structures are fully distributed.

\begin{figure}[ht]
  \hfill\hbox{\includegraphics[width=0.8\columnwidth]{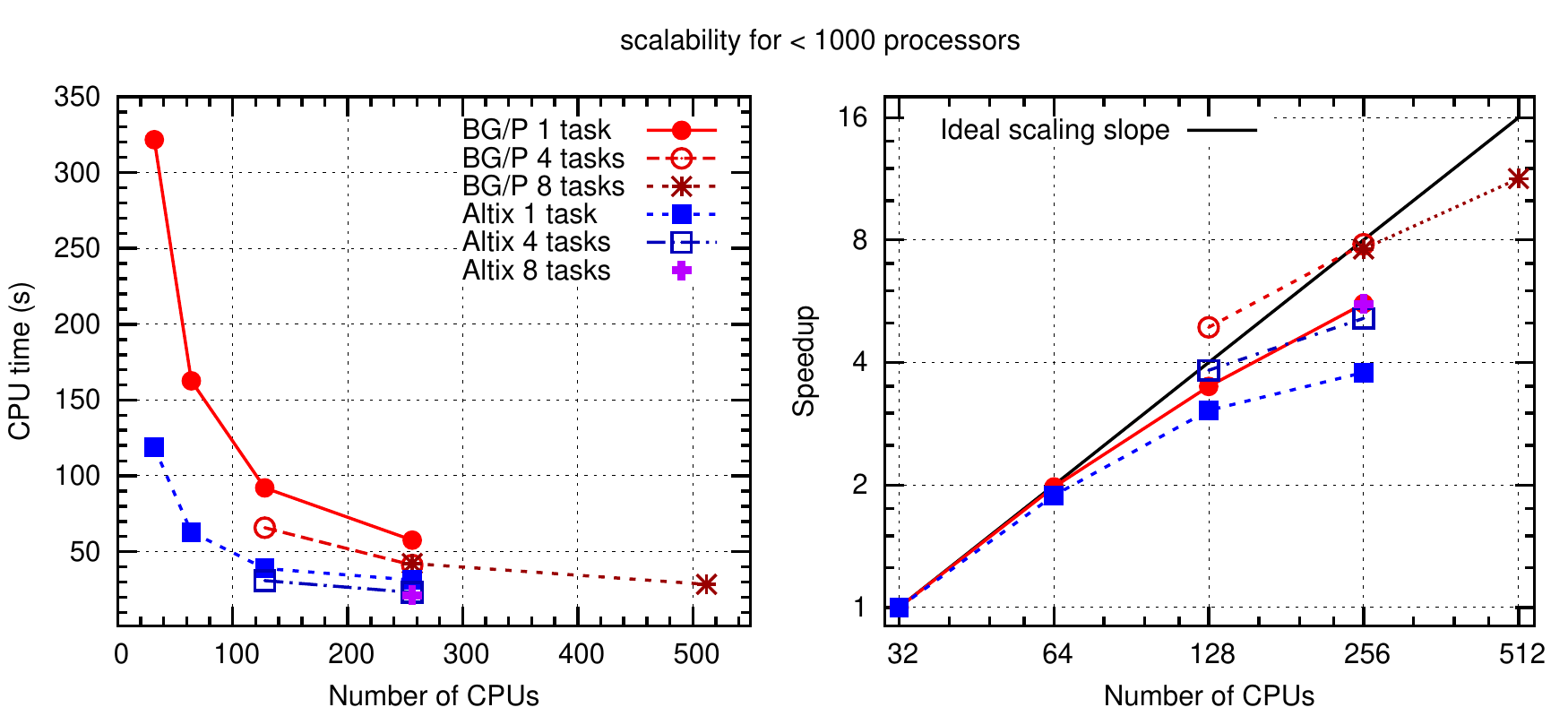}\quad}
  \caption{Scalability for medium-size calculations (\texttt{CP} code). 
   CPU time (s) per electronic time step (left panel) and
   speedup with respect to 32 processors (right panel) as a function
   of the number of processors and for different numbers $n_{task}$ 
   of task groups, on a IBM BlueGene/P (BG/P) and on a SGI Altix.
   The system is a fragment of an $A\beta-$peptide in water containing 
   838 atoms and 2311 electrons in a $22.1\times22.9\times19.9$ \AA$^3$ 
   cell, ultrasoft pseudopotentials, $\Gamma$ point, 25 Ry and 250 Ry
   cutoff for the orbitals and the charge density respectively.}
  \label{fig:scala}
\end{figure}

This being said, the size and nature of the specific application set
quite natural limits to the maximum number of processors up to which
the performances of the various codes are expected to scale. For
instance, the number of ${\bf k}-$points calculation sets a natural
limit to the size of each pool, or the number of electronic bands sets
a limit for the parallelization of the linear algebra
operations. Moreover some numerical algorithms scale better than
others. For example, the use of norm-conserving pseudopotentials
allows for a better scaling than ultrasoft pseudopotentials for a same
system, because a larger plane-wave basis set and a larger real- and
reciprocal-space grids are required in the former case.  On the other
hand, using ultrasoft pseudopotentials is generally faster because the
use of a smaller basis set is obviously more efficient, even though
the overall parallel performance may not be as good.

\begin{figure}[ht]
  \hfill\hbox{\includegraphics[width=0.8\columnwidth]{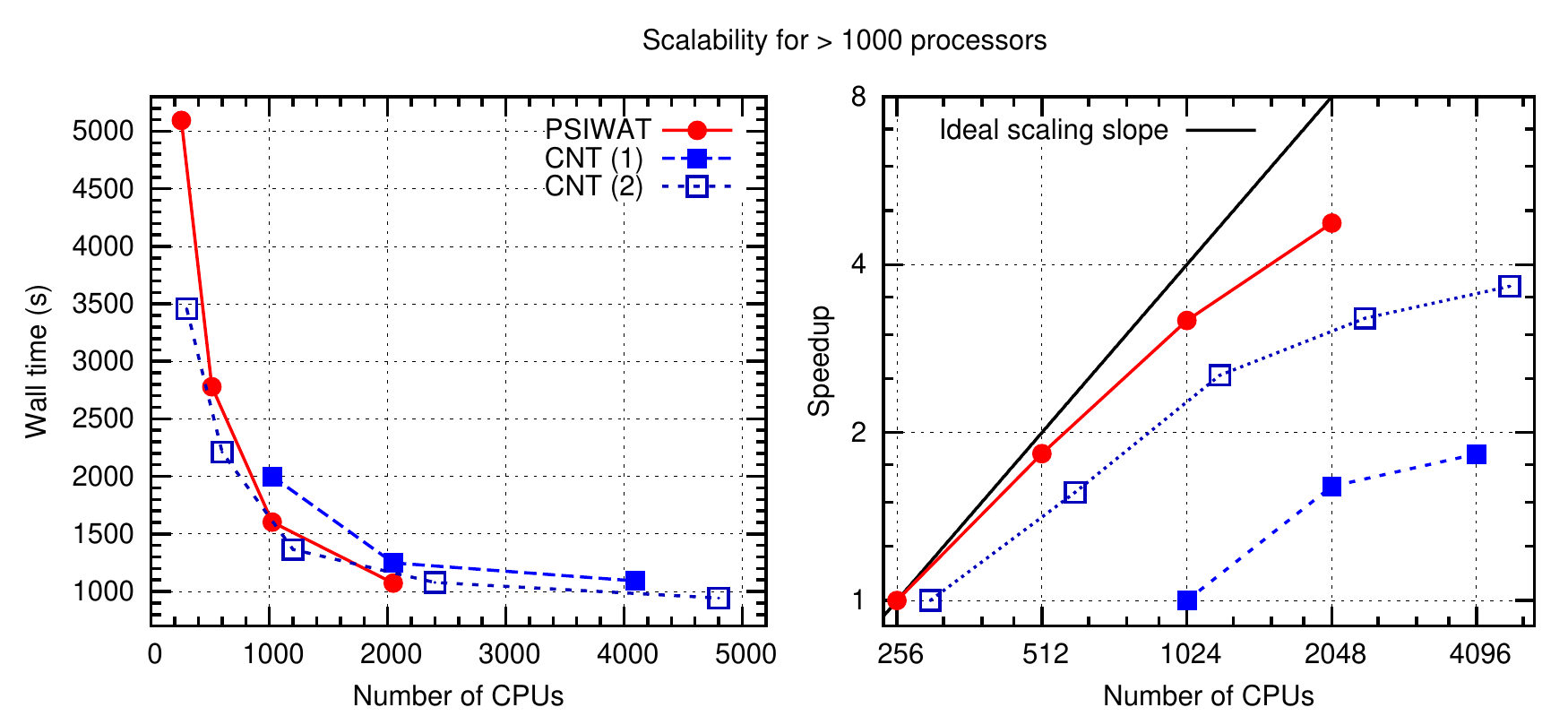}\quad}
  \caption{Scalability for large-scale calculations:
   Wall time (left panel) and speedup (right panel) as a function
   of the number of processors. PSIWAT: \texttt{PWscf} code, $n_{pool}=4$,
   $n_{task}=4$, on a Cray XT 4. The system is a gold surface covered 
   by thiols in interaction with water, 4 ${\bf k}-$points,
   $10.59 \times 20.53 \times 32.66$ \AA$^3$ cell, 587 atoms, 2552 electrons.
   CNT (1): \texttt{PWscf} code, $n_{task}=4$, on a Cray XT 4.
   The system is a porphyrin-functionalized nanotube, $\Gamma$ point,
   1532 atoms, 5232 electrons.
   CNT (2): \texttt{CP} code on a Cray XT3, same system as for CNT (1),
   Times for PSIWAT and CNT (1) are for 10 and 2 self-consistency 
   iterations, respectively; times for CNT (2) are for 10 electronic 
   steps plus 1 Car-Parrinello step, divided by 2 so that they fall
   in the same range as for CNT (1). 
   % The speedup is calculated with respect to the
   % smallest available number of processors for each set of data. 
  } 
  \label{fig:scala1}
\end{figure}

Simulations on systems containing several hundreds of atoms are by now
quite standard (see Fig. \ref{fig:scala} for an example). Scalability
does not yet extend to tens of thousands of processors as in
especially-crafted codes like \texttt{QBox} \cite{qbox}, but excellent
scalability on up to 4800 processors has been demonstrated (see Fig.
\ref{fig:scala1}) even for cases where coarse-grained parallelization
does not help, using only MPI parallelization. We remarks that the
results for CNT (2) in Fig. \ref{fig:scala1} were obtained with an
earlier version of the \texttt{CP} code that didn't use ScaLAPACK; the
current version performs better in terms of scalability.

The efforts of the \qe\ developers' team are not limited to the
performance on massively parallel architectures. Special attention
is also paid to optimize the performances for simulations of intermediate
size (on systems comprising from several tens to a few hundreds
inequivalent atoms), to be performed on medium-size clusters, readily
available to many groups \cite{carpasqua2}. In particular, the \qe\
developers' team is now working to better exploit new hardware trends,
particularly in the field of multicore architectures. The current
version implements a partial but fully functional OpenMP parallelization
\cite{OpenMP}, that is especially suitable for modern multicore CPU's. 
Mixing OpenMP with MPI also allows to extend scalability towards 
a higher number of processors, by adding a parallelization level 
on top of what can already be achieved using MPI. Preliminary tests
on realistic physical systems demonstrate scalability up to 65536 
cores, so far.

Looking ahead, future developments will likely focus on hybrid systems 
with hardware accelerators (GPUs and cell co-processors).

\section{Perspectives and Outlook}

Further developments and extensions of \qe\ will be driven by the
needs of the community using it and working on it. Many of the
soon-to-come additions will deal with excited-state calculations
within time-dependent DFT (TDDFT \cite{RG:84,TDDFT-Book}) and/or
many-body perturbation theory \cite{Onida:02}. A new approach to the
calculation of optical spectra within TDDFT has been recently
developed \cite{tddfpt}, based on a finite-frequency generalization of
density-functional perturbation theory \cite{bgt,dfpt1}, and
implemented in \qe. Another important development presently under way
is an efficient implementation of GW calculations for large systems
(whose size is of the order of a few hundreds inequivalent atoms)
\cite{Umari:arXiv0811.1453}. The implementation of efficient
algorithms for calculating correlation energies at the RPA level is
also presently under way \cite{sdg-vdw1,sdg-vdw2,viet-PhD}. It is
foreseen that by the time this paper will appear, many of these
developments will be publicly released.

It is hoped that many new functionalities will be made available to
\qe\ users by external groups who will make their own software
compatible/interfaceable with \qe. At the time of the writing of the
present paper, third-party scientific software compatible with \qe\
and available to its
users' community include: \texttt{yambo}, a general-purpose code for
excited-state calculations within many-body perturbation theory
\cite{yambo}; \texttt{casino}, a code for electronic-structure quantum
Monte Carlo simulations \cite{casino}; \texttt{want}, a code for the
simulation of ballistic transport in nanostructures, based on Wannier
functions \cite{want}; \texttt{xcrysden}, a molecular graphics
application, especially suited for periodic structures
\cite{xcrysden}. The \texttt{qe-forge} portal is expected to burst the
production and availability of third-party software compatible with
\qe. Among the projects already available, or soon-to-be available, on
\texttt{qe-forge}, we mention: \texttt{SaX} \cite{sax}, an open-source
project implementing state-of-the-art many-body perturbation theory
methods for excited states; \texttt{dmft} \cite{dmft}, a code to
perform Dynamical Mean-Field Theory calculations on top of a
tight-binding representation of the DFT band structure; \texttt{qha},
a set of codes for calculating thermal properties of materials within
the quasi-harmonic approximation \cite{qha}; \texttt{pwtk}, a fully
functional Tcl scripting interface to PWscf \cite{pwtk}.

Efforts towards better interoperability with third-party software will
be geared towards releasing accurate specifications for data
structures and data file formats and providing interfaces to and from
other codes and packages used by the scientific community.  Further
work will be also devoted to the extension to the US-PPs and PAW
schemes of the parts of \qe\ that are now limited to NC-PPs.

The increasing availability of massively parallel machines will likely
lead to an increased interest towards large-scale calculations. The
ongoing effort in this field will continue.  A special attention will
be paid to the requirements imposed by the architecture of the new
machines, in particular multicore CPUs, for which a mixed OpenMP-MPI
approach seems to be the only viable solution yielding maximum
performances. Grid computing and the commoditization of computer
cluster will also lead to great improvements in high-throughput
calculations for materials design and discovery.
 
The new trend towards distributed computing is exemplified by the
recent development of the {\em Vlab} cyber-infrastructure
(CI) \cite{vlab1,vlab2}, a service-oriented architecture (SOA) that
uses \qe\ as the back-end computational package plus a web
portal\cite{vlab3}. This SOA consists of scientific workflows for
calculations of high-pressure (P) and temperature (T) properties of
materials\cite{vlab4}, programmed as a collection of web services
running in distributed environments, plus analysis tools to monitor
workflow execution and visualization tools. High PT properties of a
inexhaustible series of \ minerals is essential for the interpretation
of seismic data and as input for geodynamic simulations. The {\em
  VLab}-CI was developed to: 1) handle the job deluge created by the
large number of points ($10^2$-$10^4$) in the parameter (pressures,
strains, phonon {\bf q}-points, composition) space sampled by these
calculations, each point consisting of a first-principle task 
(\texttt{PWscf} or \texttt{PHonon} execution); 
2) handle the information flow between
multi-leveled groups of tasks, with outputs from one level used to
generate inputs for the next level; 3) harness the scalable aggregated
throughput power of scattered computational resources.

\ack

The \qe\ project is an initiative of the CNR-INFM DEMOCRITOS National
Simulation Center in Trieste (Italy) and its partners, in
collaboration with MIT, Princeton University, the University of
Minnesota, the Ecole Polytechnique F\'ed\'erale de Lausanne, the
Universit\'e Pierre et Marie Curie in Paris, the Jo\v zef Stefan
Institute in Ljubljana, and the S3 research center in Modena. Many of
the ideas embodied in the \qe\ codes have flourished in the very
stimulating environment of the International School of Advanced
Studies, SISSA, where Democritos is hosted, and have benefited from
the ingenuity of generations of graduate students and young postdocs.  SISSA
and the CINECA National Supercomputing Center are currently providing
valuable support to the \qe\ project.

\appendix
\section*{Appendix}
\setcounter{section}{1}

This appendix contains the description of some algorithms used in \qe\
that have not been documented elsewhere.

\subsection{Self-consistency}

The problem of finding a self-consistent solution to the KS
equations can be recast into the solution of a nonlinear problem
\begin{equation}
  {\bf x} = F[{\bf x}], \qquad {\bf x} = (x_1,x_2,\ldots,x_N),
\end{equation}
where vector ${\bf x}$ contains the $N$ Fourier components or
real-space values of the charge density $\rho$ or the KS
potential $V$ (the sum of Hartree and exchange-correlation
potentials); $F[{\bf x}^{(in)}]$ is a functional of the input charge
density or potential ${\bf x}^{(in)}$, yielding the output vector
${\bf x}^{(out)}$ via the solution of KS equations. A solution
can be found via an iterative procedure.  \texttt{PWscf} uses an
algorithm based on the modified Broyden method \cite{scf} in which
${\bf x}$ contains the components of the charge density in reciprocal
space. Mixing algorithms typically find the optimal linear combination
of a few ${\bf x}^{(in)}$ from previous iterations, that minimizes
some suitably defined norm $||{\bf x}^{(out)}-{\bf x}^{(in)}||$,
vanishing at convergence, that we will call in the following ``scf
norm''.

Ideally, the scf norm is a measure of the self-consistency error on
the total energy.  Let us write an estimate of the latter for the
simplest case: an insulator with NC PPs and simple LDA or GGA.  At a
given iteration we have
\begin{equation}
  \left(-{\hbar^2\over 2m}\nabla^2+V_{ext}({\bf r})+V^{(in)}({\bf
      r})\right) \psi_i({\bf r}) = \epsilon_i \psi_i({\bf r}),
\end{equation}
where $\epsilon_i$ and $\psi_i$ are KS energies and orbitals
respectively, $i$ labels the occupied states, $V_{ext}$ is the sum of
the PPs of atomic cores (written for simplicity as a local potential),
the input Hartree and exchange-correlation potential $V^{(in)}({\bf
  r}) = V_{Hxc}[\rho^{(in)}({\bf r})]$ is a functional of the input
charge density $\rho^{(in)}$.  The output charge density is given by
\begin{equation}
\rho^{(out)}({\bf r})= \sum_i |\psi_i({\bf r})|^2. 
\end{equation}
Let us compare the DFT energy calculated in the standard way:
\begin{equation}
  E  =  \sum_i \int\psi^*_i({\bf r}) \left( -{\hbar^2\over
      2m}\nabla^2 + V_{ext}({\bf r}) \right) \psi_i({\bf r})d{\bf
    r} + E_{Hxc}[\rho^{(out)}],
\end{equation}
%  $$\displaylines{
%   \quad E  =  \sum_i \int\psi^*_i({\bf r}) \left( -{\hbar^2\over
%       2m}\nabla^2 + V_{ext}({\bf r}) \right) \psi_i({\bf r})d{\bf
%     r}  \hfill\cr\hfill + E_{Hxc}[\rho^{(out)}], \quad (\theequation)
% }$$
where $E_{Hxc}$ is the Hartree and exchange-correlation energy, with
the Harris-Weinert-Foulkes functional form, which doesn't use
$\rho^{(out)}$:
$$
\displaylines{
  \quad  E' =
  \sum_i \int\psi^*_i({\bf r}) 
  \left( 
    -{\hbar^2\over 2m} \nabla^2 + V_{ext}({\bf r}) + V^{(in)}({\bf r})
  \right)
  \psi_i({\bf r})d{\bf r}  \cr 
  \hfill - \int\rho^{(in)}V^{(in)}({\bf r})
  + E_{Hxc}[\rho^{(in)}]
  \quad\stepcounter{equation}(\theequation)
}
$$
Both forms are variational, i.e. the first-order variation of the
energy with respect to the charge density vanish, and both converge to
the same result when self-consistency is achieved.  Their difference
can be approximated by the following expression, in which only the
dominant Hartree term is considered:
\begin{eqnarray}
  E-E'& \simeq& {1\over2} \int {\Delta\rho({\bf r }) \Delta\rho({\bf
      r'}) \over  |{\bf r}-{\bf r}'|} d{\bf r} d{\bf r'} \cr & = &
           {1\over2} \int \Delta\rho({\bf r }) \Delta V_H({\bf r'})
           d{\bf r} 
\end{eqnarray}
where $\Delta\rho = \rho^{(out)}-\rho^{(in)}$ and $\Delta V_H$ is the
Hartree potential generated by $\Delta\rho$.  Moreover it can be shown
that, when exchange and correlation contributions to the electronic
screening do not dominate over the electrostatic ones, this quantity
is an upper bound to the self-consistent error incurred when using the
standard form for the DFT energy.  We therefore take this term, which
can be trivially calculated in reciprocal space, as our squared scf
norm:
\begin{equation}
  ||\rho^{(out)}-\rho^{(in)}||^2 = {4\pi e^2\over \Omega} \sum_{\bf G}
  {|\Delta\rho({\bf G})|^2 \over G^2},
\end{equation}
where ${\bf G}$ are the vectors in reciprocal space and $\Omega$ is the
volume of the unit cell.

Once the optimal linear combination of $\rho^{(in)}$ from previous
iterations (typically 4 to 8) is determined, one adds a step in the
new search direction that is, in the simplest case, a fraction of the
optimal $\Delta \rho$ or, taking advantage of some approximate
electronic screening\cite{TFmixing}, a preconditioned $\Delta
\rho$. In particular, the simple, Thomas-Fermi, and local Thomas-Fermi
mixing described in Ref. \cite{TFmixing} are implemented and used.

The above algorithm has been extended to more sophisticated
calculations, in which the ${\bf x}$ vector introduced above may
contain additional quantities: for DFT+U, occupancies of atomic
correlated states; for meta-GGA, kinetic energy density; for PAW, the
quantities $\sum_i \langle
\psi_i|\beta_n\rangle\langle\beta_m|\psi_i\rangle$, where the $\beta$
functions are the atomic-based projectors appearing in the PAW
formalism. The scf norm is modified accordingly in such a way to
include the additional variables in the estimated self-consistency
error.

\subsection{Iterative diagonalization}

During self-consistency one has to solve the generalized eigenvalue
problem for all $N$ occupied states
\begin{equation}
  H\psi_i = \epsilon_i S\psi_i,\qquad i=1,\ldots,N
  \label{diag}
\end{equation}
in which both $H$ (the Hamiltonian) and $S$ (the overlap matrix) are
available as operators (i.e. $H\psi$ and $S\psi$ products can be
calculated for a generic state $\psi$ ).  Eigenvectors are normalized
according to the generalized orthonormality constraints $\langle\psi_i
|S|\psi_j\rangle=\delta_{ij}$.  This problem is solved using iterative
methods. Currently \texttt{PWscf} implements a block Davidson
algorithm and an alternative algorithm based on band-by-band
minimization using conjugate gradient.

\subsubsection{Davidson}

One starts from an initial set of orthonormalized trial orbitals
$\psi_i^{(0)}$ and of trial eigenvalues
$\epsilon_i^{(0)}=\langle\psi^{(0)}_i |H|\psi^{(0)}_i\rangle$.  The
starting set is typically obtained from the previous scf iteration, if
available, and if not, from the previous time step, or optimization
step, or from a superposition of atomic orbitals. We introduce the
{\em residual vectors}
\begin{equation}
  g^{(0)}_i=(H-\epsilon_i^{(0)}S)\psi^{(0)}_i,
\end{equation}
a measure of the error on the trial solution, and the {\em correction
  vectors} $\delta\psi_i^{(0)} = Dg^{(0)}_i$, where $D$ is a suitable
approximation to $(H-\epsilon^{(0)}_iS)^{-1}$.  The eigenvalue problem
is then solved in the $2N$-dimensional subspace spanned by the {\em
  reduced basis set} $\phi^{(0)}$, formed by $\phi^{(0)}_i =
\psi^{(0)}_i$ and $ \phi^{(0)}_{i+N} = \delta\psi^{(0)}_i$:
\begin{equation}
  \sum_{k=1}^{2N} (H_{jk} - \epsilon_i S_{jk})c^{(i)}_k = 0,
\end{equation}
where 
\begin{equation}
  H_{jk} = \langle\phi^{(0)}_j |H|\phi^{(0)}_k\rangle,\quad
  S_{jk} = \langle\phi^{(0)}_j |S|\phi^{(0)}_k\rangle.
\end{equation}
Conventional algorithms for matrix diagonalization are used in this
step.  A new set of trial eigenvectors and eigenvalues is obtained:
\begin{equation}
  \psi_i^{(1)}=\sum_{j=1}^{2N} c^{(i)}_j\phi^{(0)}_j,\quad
  \epsilon_i^{(1)}=\langle\psi^{(1)}_i |H|\psi^{(1)}_i\rangle
\end{equation}
and the procedure is iterated until a satisfactory convergence is
achieved. Alternatively, one may enlarge the reduced basis set with
the new correction vectors $\delta\psi_i^{(1)} = Dg^{(1)}_i$, solve a
$3N$-dimensional problem, and so on, until a prefixed size of the
reduced basis set is reached. The latter approach is typically
slightly faster at the expenses of a larger memory usage.

The operator $D$ must be easy to estimate. A natural choice in the PW
basis set is a diagonal matrix, obtained keeping only the diagonal
term of the Hamiltonian:
\begin{equation}
  \langle {\bf k}+{\bf G} | D| {\bf k}+{\bf G}'\rangle =
  {\delta_{\bf GG'}\over \langle {\bf k}+{\bf G} | H - 
    \epsilon  S | {\bf k}+{\bf G}\rangle}
\end{equation}
where ${\bf k}$ is the Bloch vector of the electronic states under
consideration, $| {\bf k}+{\bf G'}\rangle$ denotes PWs, $\epsilon$ an
estimate of the highest occupied eigenvalue.  Since the Hamiltonian is
a diagonally dominant operator and the kinetic energy of PWs is the
dominant part at high ${\bf G}$, this simple form is very effective.

\subsubsection{Conjugate-Gradient}

The eigenvalue problem of Eq.(\ref{diag}) can be recast into a
sequence of constrained minimization problems:
\begin{equation}
  \mbox{min}\left[\langle\psi_i|H | \psi_i\rangle - \sum_{j\le i}
    \lambda_j\left(\langle\psi_i|S|\psi_j\rangle-\delta_{ij}\right)
  \right],
\end{equation}
where the $\lambda_j$ are Lagrange multipliers.  This can be solved
using a preconditioned conjugate gradient algorithm with minor
modifications to ensure constraint enforcement. The algorithm here
described was inspired by the conjugate-gradient algorithm of
Ref. \cite{arias}, and is similar to one of the variants described in
Ref. \cite{cgdiago}.

Let us assume that eigenvectors $\psi_j$ up to $j=i-1$ have already
been calculated. We start from an initial guess $\psi^{(0)}$ for the
$i$-th eigenvector, such that $\langle \psi^{(0)} | S |
\psi^{(0)}\rangle=1$ and $\langle \psi^{(0)} | S | \psi_j\rangle=0$.
We introduce a diagonal precondition matrix $P$ and auxiliary
functions $y = P^{-1}\psi $ and solve the equivalent problem
\begin{equation}
  \mbox{min}\left[\langle y|\widetilde H |y\rangle - 
    \lambda \left(\langle y |\widetilde S|y\rangle-1\right)
  \right],
  \label{min}
\end{equation}
where $\widetilde H = PHP$, $\widetilde S = PSP$, under the additional
orthonormality constraints $ \langle y|PS|\psi_j\rangle = 0$.  The
starting gradient of Eq.(\ref{min})) is given by
\begin{equation}
  g^{(0)} = (\widetilde H - \lambda \widetilde S) y^{(0)}.
\end{equation}
By imposing that the gradient is orthonormal to the starting vector:
$\langle g^{(0)} | \widetilde S|y^{(0)}\rangle=0$, one determines the
value of the Lagrange multiplier:
\begin{equation}
  \lambda ={\langle y^{(0)} | \widetilde S\widetilde H|y^{(0)}\rangle
    \over \langle y^{(0)} | \widetilde S^2|y^{(0)}\rangle}.
\end{equation}
The remaining orthonormality constraints are imposed on $Pg^{(0)}$ by
explicit orthonormalization (e.g. Gram-Schmid) to the $\psi_j$.  We
introduce the {\em conjugate gradient} $h^{(0)}$, which for the first
step is set equal to $g^{(0)}$ (after orthonormalization), and the
normalized direction $n^{(0)} = h^{(0)}/\langle h^{(0)} | \widetilde S
|h^{(0)}\rangle^{1/2}$.  We search for the minimum of $\langle y^{(1)}
| \widetilde H|y^{(1)}\rangle$ along the direction $y^{(1)}$, defined
as:~\cite{arias}
\begin{equation}
  y^{(1)}= y^{(0)}\cos\theta + n^{(0)}\sin\theta.
\end{equation}
This form ensures that the constraint on the norm is correctly
enforced. The calculation of the minimum can be analytically performed
and yields
\begin{equation}
  \theta= {1\over 2} \mbox{atan}\left( {a^{(0)}\over \epsilon^{(0)}-b^{(0)}}
  \right),
\end{equation}
where $a^{(0)}=2\Re \langle y^{(0)} | \widetilde H| n^{(0)}\rangle$,
$b^{(0)}=\langle n^{(0)} | \widetilde H| n^{(0)}\rangle$, and
$\epsilon^{(0)}=\langle y^{(0)} | \widetilde H|y^{(0)}\rangle$.  The
procedure is then iterated; at each step the conjugate gradient is
calculated from the gradient and the conjugate gradient at the
previous step, using the Polak-Ribi\`ere formula:
\begin{eqnarray}
  h^{(n)} & =&  g^{(n)}  + \gamma^{(n-1)} h^{(n-1)} , \\
  \gamma^{(n-1)} &=& {\langle g^{(n)}-g^{(n-1)} |
    \widetilde S| g^{(n)}\rangle 
    \over \langle g^{(n-1)} | \widetilde S| g^{(n-1)}\rangle }.
\end{eqnarray}
$h^{(n)}$ is subsequently re-orthogonalized to $y^{(n)}$.  We remarks
that in the practical implementation only $Pg$ and $Ph$ need to be
calculated and that only $P^2$ -- the analogous of the $D$ matrix of
Davidson algorithm -- is actually used. A kinetic-only form of $P^2$
has proved satisfactory:
\begin{equation}
  \langle {\bf k}+{\bf G} | P^2| {\bf k}+{\bf G}'\rangle =
  {2m \over \hbar^2 ({\bf k}+{\bf G})^2} 
  \delta_{\bf GG'}.
\end{equation}

\subsection{Wavefunction extrapolation}

In molecular dynamics runs and in structural relaxations,
extrapolations are employed to generate good initial guesses for the
wavefunctions at time $t+dt$ from wavefunctions at previous time
steps. The extrapolation algorithms used are similar to those
described in Ref. \cite{arias}. The alignment procedure, needed when
wavefunctions are the results of a self-consistent calculation, is as
follows. The overlap matrix $O_{ij}$ between wavefunctions at
consecutive time steps:
\begin{equation}
  O_{ij} = \langle\psi_i(t+dt) | S (t+dt)| \psi_j(t)\rangle,
\end{equation}
can be used to generate the unitary transformation $U$ \cite{mead}
that aligns $\psi(t+dt)$ to $\psi(t)$: $ \psi^{\parallel}_i(t+dt)
=\sum_j U_{ij} \psi_j(t+dt) $.  Since $O$ is not unitary, it needs to
be made unitary via e.g. the unitarization procedure
\begin{equation}
  \quad U=(O^\dagger O)^{-1/2} O^\dagger.
\end{equation}
The operation above is performed using a singular value decomposition:
let the overlap matrix be $O=vDw$, where $v$ and $w$ are unitary
matrix and $D$ is a diagonal non-negative definite matrix, whose
eigenvalues are close to 1 if the two sets of wavefunctions are very
similar. The needed unitary transformation is then simply given by $U
\simeq w^\dagger v^\dagger$. This procedure is simpler than the
original proposal and prevents the alignment algorithm to break in the
occasional situation where, due to level crossing in the band
structure between subsequent time steps, one or more of the
eigenvalues of the $D$ matrix vanish.

\subsection{Symmetry}

Symmetry is exploited almost everywhere, with the notable exception of
\texttt{CP}. The latter is devised to study aperiodic systems or large
supercells where symmetry is either absent or of little use even if
present.

In addition to lattice translations, the space group of a crystal
contains symmetry operations $\hat S$ combining rotations and
translations that leave the crystal unchanged: $\hat S \equiv \{
R|{\bf f}\}$, where $R$ is a $3\times 3$ orthogonal matrix, ${\bf f}$
is a vector (called fractional translation) and symmetry requires that
any atomic position, $\tau_s$ is transformed into an equivalent one,
$\hat S \tau_s \equiv R (\tau_s + {\bf f}) = \tau_{\hat S(s)} $.  The
rotational part of these operations defines the crystal point group.

As a consequence of symmetry, roto-translated KS orbitals are
KS orbitals with the rotated Bloch vector: $\hat S\psi_{i,\bf
  k}({\bf r}) \equiv \psi_{i,\bf k}(R^{-1}{\bf r} -{\bf f}) = \psi_{i,
  R\bf k}({\bf r})$. Where, strictly speaking, the resulting
wave-function at $R\bf k$ does not necessarily have the same band
index as the original one but could be some unitary transformation of
states at $R\bf k$ that share with it the same single-particle
eigenvalue.  Since quantities of physical interest are invariant for
unitary rotations among degenerate states this additional complication
has no effect on the final result.

This is the basis for the {\em symmetrization} procedure used in
\texttt{PWscf}.  One introduces a non-symmetrized charge density
(labeled by superscript $^{(ns)}$) calculated on the irreducible BZ
(IBZ):
\begin{equation}
  \rho^{(ns)}({\bf r}) =
  \sum_i\sum_{{\bf k}\in \mathrm{IBZ}} w_{\bf k} |\psi_{i,\bf k}({\bf r})|^2.
\end{equation}
The factors $w_{\bf k}$ (``weights'') are proportional to the number
of vectors in the star (i.e. inequivalent ${\bf k}$ vectors among all
the $\{R{\bf k}\}$ vectors generated by the point-group rotations) and
are normalized to 1: $\sum_{{\bf k}\in \mathrm{IBZ}}w_{\bf
  k}=1$. Weights can either be calculated or deduced from the
literature on the special-point
technique\cite{ChadiCohen,MonkhorstPack}.  The charge density is then
symmetrized as:
\begin{equation}
\rho({\bf r})= {1\over N_s} \sum_{\hat S} \hat S \rho^{(ns)}({\bf r}) 
  = {1\over N_s} \sum_{ \hat S} \rho^{(ns)}(R^{-1}{\bf r}-{\bf f})
\end{equation}
where the sum runs over all $N_s$ symmetry operations. 

The symmetrization technique can be extended to all quantities that
are expressed as sums over the BZ.  Hellmann-Feynman forces ${\bf
  F}_s$ on atom $s$ are thus calculated as follows:
\begin{equation}
{\bf F}_s = {1\over N_s}\sum_{\hat S} \hat S {\bf F}^{(ns)}_s
  = {1\over N_s}\sum_{\hat S} R {\bf F}^{(ns)}_{\hat S^{-1}(s)},
\end{equation}
where $\hat S^{-1}(s)$ labels the atom into which the $s-$th atom
transforms (modulo a lattice translation vector) after application of
$\hat S^{-1}$, the symmetry operation inverse of $\hat S$.  In a
similar way one determines the symmetrized stress, using the rule for
matrix transformation under a rotation:
\begin{equation}
\sigma_{\alpha\beta} =
  {1\over N_s}\sum_{\hat S} \sum_{\gamma,\delta=1}^3
     R_{\alpha\gamma}R_{\beta\delta} \; \sigma^{(ns)}_{\gamma\delta}.
\end{equation}

The \texttt{PHonon} package supplements the above technique with a
further strategy. Given the phonon wave-vector ${\bf q}$, the small
group of ${\bf q}$ (the subgroup $\hat S_{\bf q}$ of crystal symmetry
operations that leave ${\bf q}$ invariant) is identified and the
reducible representation defined by the $3N_{at}$ atomic displacements
along cartesian axis is decomposed into $n_{irr}$ irreducible
representations (irreps) $\gamma^{({\bf q})}_j,
j=1,\ldots,n_{irr}$. The dimensions of the irreducible representations
are small, with $\nu_j\le 3$ in most cases, up to 6 in some special
cases (zone-boundary wave-vectors $\bf q$ in nonsymmorphic
groups). Each irrep, $j$, is therefore defined by a set of $\nu_j$
linear combinations of atomic displacements that transform into each
other under the symmetry operations of the small group of ${\bf
  q}$. In the self-consistent solution of the linear response
equations, only perturbations associated to a given irrep need to be
treated together and different irreps can be solved
independently. This feature is exploited to reduce the amount of
memory required by the calculation and is suitable for coarse-grained
parallelization and for execution on a Grid
infrastructure~\cite{grid}.

The wavefunction response, $\Delta \psi^{(j,\alpha)}_{{\bf k}+{\bf
    q},i}({\bf r})$, to displacements along irrep $j$, $\gamma^{({\bf
    q})}_{j,\alpha}$ (where $\alpha=1,\ldots,\nu_j$ labels different
partners of the given irrep), is then calculated.  The
lattice-periodic unsymmetrized charge response, $\Delta \rho_{{\bf
    q},j,\alpha}^{(ns)}({\bf r})$, has the form:
\begin{equation}
  \Delta \rho_{{\bf q},j,\alpha}^{(ns)}({\bf r}) = e^{-i{\bf
      q}\cdot{\bf r}} 4 \sum_i \sum_{{\bf k}\in \mathrm{IBZ}({\bf q})}
  w_{\bf k}\psi^*_{{\bf k},i}({\bf r}) \Delta\psi^{(j,\alpha)}_{{\bf
      k}+{\bf q},i}({\bf r}), 
\end{equation}
where the notation IBZ$({\bf q})$ indicates the IBZ calculated
assuming the small group of {\bf q} as symmetry group, and the weights
$w_{\bf k}$ are calculated accordingly.  The symmetrized charge
response is calculated as
\begin{equation}
  \Delta \rho_{{\bf q},j,\alpha}({\bf r}) = {1\over N_s({\bf q})} 
  \sum_{\hat S_{\bf q}} e^{-i{\bf qf}} \sum_{\beta=1}^{\nu_j} D(\hat
  S_q)_{\beta \alpha}  \;  \Delta \rho_{{\bf
      q},j,\beta}^{(ns)}(R^{-1}{\bf r} -{\bf f} ) 
\end{equation}
where $D(\hat S_q)$ is the matrix representation of the action of the
symmetry operation $\hat S_{\bf q} \equiv \{R|{\bf f}\}$ for the
$j-$th irrep $\gamma^{({\bf q})}_{j}$. At the end of the
self-consistent procedure, the force constant matrix
$C_{s\alpha,t\beta}({\bf q})$ (where $s,t$ label atoms, $\alpha,
\beta$ cartesian coordinates) is calculated.  Force constants at all
vectors in the star of ${\bf q}$ are then obtained using symmetry:
\begin{equation}
  C_{s\alpha,t\beta}( R {\bf q}) = 
  \sum_{\gamma,\delta} R_{\alpha\delta} R_{\beta\gamma}
  C_{\hat S^{-1}(s)\delta, \hat S^{-1}(t)\gamma}({\bf q}), 
\end{equation}
where $\hat S\equiv \{ R| {\bf f}\}$ is a symmetry operation of the
crystal group but not of the small group of ${\bf q}$.

\subsection{Fock exchange}
\newcommand{\bfr}{\mathbf{r}} \newcommand{\bfrp}{\mathbf{r'}}
\newcommand{\bk}{\mathbf{k}} \newcommand{\bkp}{\mathbf{k'}}
\newcommand{\bq}{\mathbf{q}}

Hybrid functionals are characterized by the inclusion of a fraction of
{\em exact} ({\em i.e.} non-local) Fock exchange in the definition of
the exchange-correlation functional.  For a periodic system, the Fock
exchange energy per unit cell is given by:
\begin{equation}
E_{x} = - \frac{e^2}{N} \sum_{^{\bk v}_{\bkp  v'}}
                         \int \frac{\psi_{\bk v}^*(\bfr)\psi_{\bkp
                             v'}(\bfr) \psi_{\bkp v'}^*(\bfrp)\psi_{\bk
                             v}(\bfrp) }  {|\bfr-\bfrp|}  d\bfr d\bfrp,
\end{equation}
where an insulating and non magnetic system is assumed for simplicity.
Integrals and wave-function normalizations are defined over the whole
crystal volume, $V=N\Omega$ ($\Omega$ being the unit cell volume), and
the summations run over all occupied bands and all $N$ $\bk$-points
defined in the BZ by Born-von K\'arm\'an boundary conditions.  The
calculation of this term is performed exploiting the dual-space
formalism: auxiliary codensities, $\rho_{^{\bkp, v'}_{\bk, v}}(\bfr) =
\psi^*_{\bkp, v'}(\bfr) \psi_{\bk, v}(\bfr)$ are computed in real space
and transformed to reciprocal space by FFT, where the associated
electrostatic energies are accumulated. The application of the Fock
exchange operator to a wavefunction involves additional FFTs and
real-space array multiplications. These basic operations need to be
repeated for all the occupied bands and all the points in the BZ
grid. For this reason the computational cost of the exact exchange
calculation is very high, at least an order of magnitude larger than
for non-hybrid functional calculations.

In order to limit the computational cost, an auxiliary grid of
$\bq$-points in the BZ, centered at the $\Gamma$ point, can be
introduced and the summation over $\bkp$ be limited to the subset
$\bkp = \bk + \bq$.  Of course convergence with respect to this
additional parameter needs to be checked, but often a grid coarser
than the one used for computing densities and potentials is
sufficient.

The direct evaluation of the Fock energy on regular grids in the BZ is
however problematic due to an integrable divergence that appears in
the $\bq \rightarrow 0$ limit. This problem is addressed resorting to
a procedure, first proposed by Gygi and Baldereschi \cite{hf1}, where
an integrable term that displays the same divergence is subtracted
from the expression for the exchange energy and its analytic integral
over the BZ is separately added back to it.  Some care must still be
paid \cite{sdg-vdw1} in order to estimate the contribution of the $\bq
= 0$ term in the sum, which contains a $0/0$ limit that cannot be
calculated from information at $\bq = 0$ only. This term is estimated
\cite{sdg-vdw1} assuming that the grid of $\bq$-points used for
evaluating the exchange integrals is dense enough that a coarser grid,
including only every second point in each direction, would also be
equally accurate. Since the limiting term contributes to the integral
with different weights in the two grids, one can extract its value
from the condition that the two integral give the same result.  This
procedure removes an error proportional to the inverse of the unit
cell volume $\Omega$ that would otherwise appear if this term were
simply neglected.

\section*{References}
%% \References

\bibliographystyle{unsrt}
\bibliography{qe}% Produces the bibliography via BibTeX

\begin{thebibliography}{100}

\bibitem{marzari-2006}
N.~Marzari.
\newblock {\em MRS BULLETIN}, 31:681, 2006.

\bibitem{dft1}
R.~G. Parr and W.~Yang.
\newblock {\em Density Functional Theory of Atoms and Molecules}.
\newblock Oxford University Press, 1989.

\bibitem{dft2}
R.~M. Dreizler and E.~K.~U. Gross.
\newblock {\em Density Functional Theory}.
\newblock Springer-Verlag, 1990.

\bibitem{martin-book}
R.M. Martin.
\newblock Cambridge University Press, Cambridge, UK, 2004.

\bibitem{adf}
E.~J. Baerends, J.~Autschbach, A.~B\'erces, F.M. Bickelhaupt, C.~Bo, P.~M.
  Boerrigter, L.~Cavallo, D.~P. Chong, L.~Deng, R.~M. Dickson, D.~E. Ellis,
  M.~van Faassen, L.~Fan, T.~H. Fischer, C.~Fonseca Guerra, S.~J.~A. van
  Gisbergen, A.~W. G\"otz, J.~A. Groeneveld, O.~V. Gritsenko, M.~Gr\"uning,
  F.~E. Harris, P.~van~den Hoek, C.~R. Jacob, H.~Jacobsen, L.~Jensen, G.~van
  Kessel, F.~Kootstra, M.~V. Krykunov, E.~van Lenthe, D.~A. McCormack,
  A.~Michalak, J.~Neugebauer, V.~P. Nicu, V.~P. Osinga, S.~Patchkovskii,
  P.~H.~T. Philipsen, D.~Post, C.~C. Pye, W.~Ravenek, J.~I. Rodriguez, P.~Ros,
  P.~R.~T. Schipper, G.~Schreckenbach, J.~G. Snijders, M.~Sol\`a, M.~Swart,
  D.~Swerhone, G.~te~Velde, P.~Vernooijs, L.~Versluis, L.~Visscher, O.~Visser,
  F.~Wang, T.~A. Wesolowski, E.~M. van Wezenbeek, G.~Wiesenekker, S.~K. Wolff,
  T.~K. Woo, A.~L. Yakovlev, and T.~Ziegler.
\newblock \uppercase{A}DF2008.01, SCM, Theoretical Chemistry, Vrije
  Universiteit, Amsterdam, The Netherlands, http://www.scm.com.

\bibitem{crystal}
Roberto Dovesi, Bartolomeo Civalleri, Roberto Orlando, Carla Roetti, and
  Victor~R. Saunders.
\newblock in {\em \uppercase{R}eviews in Computational Chemistry}, Ch. 1, Vol.
  21, K. B. Lipkowitz, R. Larter, T. R. Cundari editors, John Wiley and Sons,
  New York (2005).

\bibitem{molcas}
G.~Karlstr\"om, R.~Lindh, P.-\AA. Malmqvist, B.~O. Roos, U.~Ryde, V.~Veryazov,
  P.-O. Widmark, M.~Cossi, B.~Schimmelpfennig, P.~Neogrady, and L.~Seijo.
\newblock {\em Comp. Mater. Sci.}, 28:222, 2003.

\bibitem{turbomole}
Reinhart Ahlrichs, Filipp Furche, Christof H\"attig, Willem~M. Klopper, Marek
  Sierka, and Florian Weigend.
\newblock \uppercase{T}URBOMOLE -- http://www.turbomole-gmbh.com.

\bibitem{g03}
M.~J. Frisch, G.~W. Trucks, H.~B. Schlegel, G.~E. Scuseria, M.~A. Robb, J.~R.
  Cheeseman, J.~A. Montgomery, Jr., T.~Vreven, K.~N. Kudin, J.~C. Burant, J.~M.
  Millam, S.~S. Iyengar, J.~Tomasi, V.~Barone, B.~Mennucci, M.~Cossi,
  G.~Scalmani, N.~Rega, G.~A. Petersson, H.~Nakatsuji, M.~Hada, M.~Ehara,
  K.~Toyota, R.~Fukuda, J.~Hasegawa, M.~Ishida, T.~Nakajima, Y.~Honda,
  O.~Kitao, H.~Nakai, M.~Klene, X.~Li, J.~E. Knox, H.~P. Hratchian, J.~B.
  Cross, V.~Bakken, C.~Adamo, J.~Jaramillo, R.~Gomperts, R.~E. Stratmann,
  O.~Yazyev, A.~J. Austin, R.~Cammi, C.~Pomelli, J.~W. Ochterski, P.~Y. Ayala,
  K.~Morokuma, G.~A. Voth, P.~Salvador, J.~J. Dannenberg, V.~G. Zakrzewski,
  S.~Dapprich, A.~D. Daniels, M.~C. Strain, O.~Farkas, D.~K. Malick, A.~D.
  Rabuck, K.~Raghavachari, J.~B. Foresman, J.~V. Ortiz, Q.~Cui, A.~G. Baboul,
  S.~Clifford, J.~Cioslowski, B.~B. Stefanov, G.~Liu, A.~Liashenko, P.~Piskorz,
  I.~Komaromi, R.~L. Martin, D.~J. Fox, T.~Keith, M.~A. Al-Laham, C.~Y. Peng,
  A.~Nanayakkara, M.~Challacombe, P.~M.~W. Gill, B.~Johnson, W.~Chen, M.~W.
  Wong, C.~Gonzalez, and J.~A. Pople.
\newblock Gaussian 03, \uppercase{R}evision \uppercase{C}.02.
\newblock \uppercase{G}aussian, Inc., Wallingford, CT, 2004,
  http://www.gaussian.com.

\bibitem{ms}
\uppercase{M}aterials Studio, http://accelrys.com/products/materials-studio.

\bibitem{schrodinger}
\uppercase{J}aguar: rapid ab initio electronic structure package,
  Schr\"{o}dinger, LLC., http://www.schrodinger.com/.

\bibitem{wavefunction}
\uppercase{S}partan, Wavefunction, Inc.,
  http://www.wavefun.com/products/spartan.html.

\bibitem{vasp}
G.~Kresse and J.~Furthmuller.
\newblock {\em Comp. Mater. Sci.}, 6(1):15--50, JUL 1996.
\newblock http://cmp.univie.ac.at/vasp.

\bibitem{castep}
http://www.castep.org,
  http://accelrys.com/products/materials-studio/modules/CASTEP.html.

\bibitem{cpmd}
The CPMD consortium page, coordinated by M. Parrinello and W. Andreoni,
  Copyright IBM Corp 1990-2008, Copyright MPI f\"ur Festk\"orperforschung
  Stuttgart 1997-2001, http://www.cpmd.org.

\bibitem{nwchem}
E.~J. Bylaska, W.~A. de~Jong, N.~Govind, K.~Kowalski, T.~P. Straatsma,
  M.~Valiev, D.~Wang, E.~Apra, T.~L. Windus, J.~Hammond, P.~Nichols, S.~Hirata,
  M.~T. Hackler, Y.~Zhao, P.-D. Fan, R.~J. Harrison, M.~Dupuis, D.~M.~A. Smith,
  J.~Nieplocha, V.~Tipparaju, M.~Krishnan, Q.~Wu, T.~Van Voorhis, A.~A. Auer,
  M.~Nooijen, E.~Brown, G.~Cisneros, G.~I. Fann, H.~Fruchtl, J.~Garza,
  K.~Hirao, R.~Kendall, J.~A. Nichols, K.~Tsemekhman, K.~Wolinski, J.~Anchell,
  D.~Bernholdt, P.~Borowski, T.~Clark, D.~Clerc, H.~Dachsel, M.~Deegan,
  K.~Dyall, D.~Elwood, E.~Glendening, M.~Gutowski, A.~Hess, J.~Jaffe,
  B.~Johnson, J.~Ju, R.~Kobayashi, R.~Kutteh, Z.~Lin, R.~Littlefield, X.~Long,
  B.~Meng, T.~Nakajima, S.~Niu, L.~Pollack, M.~Rosing, G.~Sandrone, M.~Stave,
  H.~Taylor, G.~Thomas, J.~van Lenthe, A.~Wong, , and Z.~Zhang.
\newblock Nwchem, a computational chemistry package for parallel computers,
  version 5.1.
\newblock \uppercase{P}acific Northwest National Laboratory, Richland,
  Washington 99352-0999, USA, http://www.emsl.pnl.gov/docs/nwchem/nwchem.html.

\bibitem{nwchem2}
R.~A. Kendall, E.~Apra, D.~E. Bernholdt, E.~J. Bylaska, M.~Dupuis, G.~I. Fann,
  R.~J. Harrison, J.~Ju, J.~A. Nichols, J.~Nieplocha, T.~P. Straatsma, T.~L.
  Windus, and A.~T. Wong.
\newblock High performance computational chemistry: An overview of nwchem a
  distributed parallel application.
\newblock {\em Comput. Phys. Commun.}, 128:260--283, 2000.

\bibitem{gamess1}
M.~W. Schmidt, K.~K. Baldridge, J.~A. Boatz, S.~T. Elbert, M.~S. Gordon, J.~H.
  Jensen, S.~Koseki, N.~Matsunaga, K.~A. Nguyen, S.~J. Su, T.~L. Windus,
  M.~Dupuis, and J.~A. Montgomery.
\newblock General atomic and molecular electronic structure system.
\newblock {\em J. Comput. Chem.}, 14:1347--1363, 1993.

\bibitem{gamess2}
M.~S. Gordon and M.~W. Schmidt.
\newblock Advances in electronic structure theory: Gamess a decade later.
\newblock in {\it Theory and Applications of Computational Chemistry, the first
  forty years}, C. E. Dykstra, G. Frenking, K. S. Kim, G. E. Scuseria editors,
  ch. 41, pp 1167-1189, Elsevier, Amsterdam, 2005,
  http://www.msg.chem.iastate.edu/GAMESS/GAMESS.html.

\bibitem{gpl}
http://www.gnu.org/licenses/.

\bibitem{abinit}
X.~Gonze, J.-M. Beuken, R.~Caracas, F.~Detraux, M.~Fuchs, G.-M. Rignanese,
  L.~Sindic, M.~Verstraete, G.~Zerah, F.~Jollet, M.~Torrent, A.~Roy, M.~Mikami,
  Ph. Ghosez, J.-Y. Raty, and D.~C. Allan.
\newblock First-principle computation of material properties: the abinit
  software project.
\newblock {\em Comp. Mater. Sci.}, 25:478, 2002.
\newblock http://www.abinit.org.

\bibitem{cp2k}
J.~VandeVondele, M.~Krack, F.~Mohamed, M.~Parrinello, T.~Chassaing, and
  J.~Hutter.
\newblock {\em Comput. Phys. Commun.}, 167:103, 2005.
\newblock http://cp2k.berlios.de.

\bibitem{dacapo}
S.~R. Bahn and K.~W. Jacobsen.
\newblock An object-oriented scripting interface to a legacy
  electronic-structure code.
\newblock {\em Comput. Sci. Eng.}, 4:56, 2002.
\newblock http://www.fysik.dtu.dk/campos.

\bibitem{gpaw}
J.~J. Mortensen, L.~B. Hansen, and K.~W. Jacobsen.
\newblock {\em Phys. Rev. B}, 71, 2005.
\newblock https://wiki.fysik.dtu.dk/gpaw.

\bibitem{PSI3}
T.~Daniel Crawford, C.~David Sherrill, Edward~F. Valeev, Justin~T. Fermann,
  Rollin~A. King, Matthew~L. Leininger, Shawn~T. Brown, Curtis~L. Janssen,
  Edward~T. Seidl, Joseph~P. Kenny, and Wesley~D. Allen.
\newblock {\em J. Comput. Chem.}, 28(9):1610--1616, 2007.
\newblock http://www.psicode.org.

\bibitem{wannier90}
A.~A. Mostofi, J.~R. Yates, Y.-S. Lee, I.~Souza, D.~Vanderbilt, and N.~Marzari.
\newblock Wannier90: A tool for obtaining maximally-localised wannier
  functions.
\newblock {\em Comput. Phys. Commun.}, 178:685, 2008.

\bibitem{qe-forge}
http://qe-forge.org.

\bibitem{laug}
E.~Anderson, Z.~Bai, C.~Bischof, S.~Blackford, J.~Demmel, J.~Dongarra,
  J.~Du~Croz, A.~Greenbaum, S.~Hammarling, A.~McKenney, and D.~Sorensen.
\newblock {\em {LAPACK} Users' Guide}.
\newblock Society for Industrial and Applied Mathematics, Philadelphia, PA,
  third edition, 1999.

\bibitem{FFTW05}
Matteo Frigo and Steven~G. Johnson.
\newblock The design and implementation of {FFTW3}.
\newblock {\em Proceedings of the IEEE}, 93(2):216--231, 2005.
\newblock special issue on "Program Generation, Optimization, and Platform
  Adaptation".

\bibitem{MPI}
{Message Passing Interface Forum}.
\newblock {\em Int. J. Supercomputer Applications}, 8 (3/4), 1994.

\bibitem{SF}
http://en.wikipedia.org/wiki/SourceForge.

\bibitem{pwpp}
W.~E. Pickett.
\newblock {\em Comput. Phys. Rep.}, 9:115, 1989.

\bibitem{ismaila08}
Ismaila Dabo, Boris Kozinsky, Nicholas~E. Singh-Miller, and Nicola Marzari.
\newblock {\em Phys. Rev. B}, 77:115139, 2008.

\bibitem{KB}
L.~Kleinman and D.~M. Bylander.
\newblock {\em Phys. Rev. Lett.}, 48:1425, 1982.

\bibitem{HSC}
D.~R. Hamann, M.~Schl\"uter, and C.~Chiang.
\newblock {\em Phys. Rev. Lett.}, 43:1494, 1979.

\bibitem{USPP}
D.~Vanderbilt.
\newblock {\em Phys. Rev. B}, 41:7892, 1990.

\bibitem{PAW1}
P.~E. Bl\"ochl.
\newblock {\em Phys. Rev. B}, 50(24):17953, Dec. 1994.

\bibitem{GGA}
J.~P. Perdew, J.~A. Chevary, S.~H. Vosko, K.~A. Jackson, D.~J. Singh, and
  C.~Fiolhais.
\newblock {\em Phys. Rev. B}, 46:6671, 1992.

\bibitem{TPSS}
J.~Tao, J.~P. Perdew, V.~N. Staroverov, and G.~E. Scuseria.
\newblock {\em Phys. Rev. Lett.}, 91:146401, 2003.

\bibitem{PBE0}
J.~P. Perdew, M.~Ernzerhof, and K.~Burke.
\newblock {\em J. Chem. Phys.}, 105:9982, 1996.

\bibitem{Becke1}
A.~D. Becke.
\newblock {\em J. Chem. Phys.}, 98:1372, 1993.

\bibitem{B3LYP}
P.~J. Stephens, F.~J. Devlin, C.~F. Chabalowski, and M.~J. Frisch.
\newblock {\em J. Phys. Chem.}, 98:11623, 1994.

\bibitem{KS}
W.~Kohn and L.~J. Sham.
\newblock {\em Phys. Rev.}, 140:A1133, 1965.

\bibitem{Hellmann}
H.~Hellmann.
\newblock {\em Einf\"uhrung in die Quantenchemie}.
\newblock Deutliche, 1937.

\bibitem{Feynman}
R.~P. Feynman.
\newblock {\em Phys. Rev.}, 56:340, 1939.

\bibitem{stress}
O.~H. Nielsen and R.~M. Martin.
\newblock {\em Phys. Rev. B}, 32:3780, 1985.

\bibitem{relativistic}
P.~Pyykk\"o.
\newblock {\em Chem. Rev.}, 88:563, 1988.

\bibitem{noncol1}
T.~Oda, A.~Pasquarello, and R.~Car.
\newblock {\em Phys. Rev. Lett.}, 80:3622, 1998.

\bibitem{noncol2}
R.~Gebauer and S.~Baroni.
\newblock {\em Phys. Rev. B}, 61:R6459, 2000.

\bibitem{CP}
R.~Car and M.~Parrinello.
\newblock {\em Phys. Rev. Lett.}, 55:2471, 1985.

\bibitem{BOMD}
R.~M. Wentzcovitch and J.~L. Martins.
\newblock {\em Sol. St. Commun.}, 78:831, 1991.

\bibitem{vcmd0}
M.~Parrinello and A.~Rahman.
\newblock {\em Phys. Rev. Lett.}, 45(14):1196--1199, Oct 1980.

\bibitem{vcmd2}
R.~M. Wentzcovitch.
\newblock {\em Phys. Rev. B}, 44:2358--2361, 1991.

\bibitem{bgt}
S.~Baroni, P.~Giannozzi, and A.~Testa.
\newblock {\em Phys. Rev. Lett.}, 58:1861, 1987.

\bibitem{dfpt1}
S.~Baroni, S.~de~Gironcoli, A.~{Dal Corso}, and P.~Giannozzi.
\newblock {\em Rev. Mod. Phys.}, 73:515, 2001.

\bibitem{dfpt2}
X.~Gonze.
\newblock {\em Phys. Rev. A}, 52:1096, 1995.

\bibitem{neb1}
G.~Henkelman and H.~J\'onsson.
\newblock {\em J. Chem. Phys.}, 111:7010, 1999.

\bibitem{neb2}
W.~E, W.~Ren, and E.~Vanden-Eijnden.
\newblock {\em Phys. Rev. B}, 66:052301, 2002.

\bibitem{neb3}
K.~J. Caspersen and E.~A. Carter.
\newblock {\em P. Natl. Acad. Sci. USA}, 102(19):6738--43, 2005.

\bibitem{ball1}
H.~J. Choi and J.~Ihm.
\newblock {\em Phys. Rev. B}, 59:2267, 1999.

\bibitem{MaxWan1}
N.~Marzari and D.~Vanderbilt.
\newblock {\em Phys. Rev. B}, 56:12847, 1997.

\bibitem{MaxWan2}
I.~Souza, N.~Marzari, and D.~Vanderbilt.
\newblock {\em Phys. Rev. B}, 65:035109, 2001.

\bibitem{Pickard_2001_a_gipaw_PRB}
C.~J. Pickard and F.~Mauri.
\newblock {\em Phys. Rev.}, B 63:245101, 2001.

\bibitem{Pickard_2003_a_gipaw_PRL}
C.~J. Pickard and F.~Mauri.
\newblock {\em Phys. Rev. Lett.}, 91:196401, 2003.

\bibitem{Taillefumier}
M.~Taillefumier, D.~Cabaret, A.~M. Flank, and F.~Mauri.
\newblock {\em Phys. Rev. B}, 66:195107, 2002.

\bibitem{ZK}
S.~Scandolo, P.~Giannozzi, C.~Cavazzoni, S.~de~Gironcoli, A.~Pasquarello, and
  S.~Baroni.
\newblock {\em Z. Kristallogr.}, 220:574, 2005.

\bibitem{stm}
J.~Tersoff and D.~R. Hamann.
\newblock {\em Phys. Rev. B}, 31:805--813, 1985.

\bibitem{elf}
A.~D. Becke and K.~E. Edgecombe.
\newblock {\em J. Chem. Phys.}, 92:5397--5403, 1990.

\bibitem{QC}
A.~Szabo and N.~Ostlund.
\newblock {\em Modern Quantum Chemistry}.
\newblock Dover, 1996.

\bibitem{macroscopic}
A.~Baldereschi, S.~Baroni, and R.~Resta.
\newblock {\em Phys. Rev. Lett.}, 61:734, 1988.

\bibitem{hdf}
{The HDF Group}.
\newblock http://www.hdfgroup.org.

\bibitem{netcdf}
http://www.unidata.ucar.edu/software/netcdf.

\bibitem{xml}
http://www.quantum-simulation.org.

\bibitem{iotk}
G.~Bussi.
\newblock http://www.s3.infm.it/iotk.

\bibitem{fhi98PP}
Martin Fuchs and Matthias Scheffler.
\newblock {\em Comp. Phys. Comm.}, 119(1):67, 1999.

\bibitem{VdB-uspp}
http://www.physics.rutgers.edu/~dhv/uspp/.

\bibitem{opium}
OPIUM -- pseudopotential generation project, http://opium.sourceforge.net.

\bibitem{PWscf}
A.~{Dal Corso}.
\newblock A pseudopotential plane waves program {\tt(pwscf)} and some case
  studies.
\newblock in {\it Lecture Notes in Chemistry}, Vol. 67, C. Pisani editor,
  Springer Verlag, Berlin (1996).

\bibitem{PAW2}
G.~Kresse and D.~Joubert.
\newblock {\em Phys. Rev. B}, 59(3):1758--1775, Jan 1999.

\bibitem{carpasqua1}
K.~Laasonen, A.~Pasquarello, R.~Car, C.~Lee, and D.~Vanderbilt.
\newblock {\em Phys. Rev. B}, 47:10142, 1993.

\bibitem{carpasqua2}
P.~Giannozzi, F.~de~Angelis, and R.~Car.
\newblock {\em J. Chem. Phys.}, 120:5903, 2004.

\bibitem{WhiteBird}
J.~A. White and D.~M. Bird.
\newblock {\em Phys. Rev. B}, 50:4954, 1994.

\bibitem{spinorbit1}
A.~{Dal Corso} and A.~{Mosca Conte}.
\newblock {\em Phys. Rev. B}, 71:115106, 2005.

\bibitem{spinorbit2}
Adriano {Mosca Conte}.
\newblock {\em Quantum mechanical modeling of nano magnetism: new tools based
  on Density-Functional theory with case applications to solids, surfaces,
  wires, and molecules}.
\newblock PhD thesis, SISSA/ISAS, Trieste, Italy, 2007.

\bibitem{LDA+U}
V.~I. Anisimov, J.~Zaanen, and O.~K. Andersen.
\newblock {\em Phys. Rev. B}, 44:943, 1991.

\bibitem{Ucococcioni}
M.~Cococcioni and S.~de~Gironcoli.
\newblock {\em Phys. Rev. B}, 71:035105, 2005.

\bibitem{KZK}
H.~Kwee, S.~Zhang, and H.~Krakauer.
\newblock {\em Phys. Rev. Lett.}, 100:126404, 2008.

\bibitem{scf}
D.~D. Johnson.
\newblock {\em Phys. Rev. B}, 38:12807, 1988.

\bibitem{ChadiCohen}
D.~J. Chadi and M.~L. Cohen.
\newblock {\em Phys. Rev. B}, 8:5747, 1973.

\bibitem{MonkhorstPack}
H.~J. Monkhorst and J.D. Pack.
\newblock {\em Phys. Rev. B}, 13:5188, 1976.

\bibitem{MP}
M.~Methfessel and A.~Paxton.
\newblock {\em Phys. Rev. B}, 40:3616, 1989.

\bibitem{cold}
N.~Marzari, D.~Vanderbilt, A.~de~Vita, and M.~C. Payne.
\newblock {\em Phys. Rev. Lett.}, 82:3296, 1999.

\bibitem{tetra}
P.~E. Bl\"ochl, O.~Jepsen, and O.K. Andersen.
\newblock {\em Phys. Rev. B}, 49:16223, 1994.

\bibitem{mermin}
N.~David Mermin.
\newblock {\em Phys. Rev.}, 137(5A):A1441--A1443, Mar 1965.

\bibitem{opt}
R.~Fletcher.
\newblock {\em Practical Methods of Optimization}.
\newblock John Wiley and Sons, 1987.

\bibitem{bfgs1}
S.~R. Billeter, A.~J. Turner, and W.~Thiel.
\newblock {\em Phys. Chem. Chem. Phys.}, 2:2177, 2000.

\bibitem{bfgs2}
S.~R. Billeter, A.~Curioni, and W.~Andreoni.
\newblock {\em Comp. Mater. Sci.}, 27:437, 2003.

\bibitem{metadyn}
C.~Micheletti, A.~Laio, and M.~Parrinello.
\newblock {\em Phys. Rev. Lett.}, 92:170601, 2004.

\bibitem{Verlet}
L.~Verlet.
\newblock {\em Phys. Rev.}, 159:98, 1967.

\bibitem{allentild}
M.~P. Allen and D.~J. Tildesley.
\newblock {\em Computer Simulations of Liquids}.
\newblock Clarendon Press, 1986.

\bibitem{vcmd1}
M.~Bernasconi, G.~L. Chiarotti, P.~Focher, S.~Scandolo, E.~Tosatti, and
  M.~Parrinello.
\newblock {\em J. Phys. Chem. Solids}, 56:501--505, 1995.

\bibitem{Efield}
P.~Umari and A.~Pasquarello.
\newblock {\em Phys. Rev. Lett.}, 89:157602, 2002.

\bibitem{souza-vanderbilt}
Ivo Souza, Jorge \'I\~niguez, and David Vanderbilt.
\newblock {\em Phys. Rev. Lett.}, 89(11):117602, Aug 2002.

\bibitem{kunc}
K.~Kunc and R.~Resta.
\newblock {\em Phys. Rev. Lett.}, 51:686, 1983.

\bibitem{tobik}
J.~Tobik and A.~{Dal Corso}.
\newblock {\em J. Chem. Phys.}, 120:9934, 2004.

\bibitem{damian06}
Dami\'{a}n~A. Scherlis, Jean-Luc Fattebert, Fran\c{c}ois Gygi, Matteo
  Cococcioni, and Nicola Marzari.
\newblock {\em J. Chem. Phys.}, 124:074103, 2006.

\bibitem{ismailaec09}
Ismaila Dabo, Eric Cances, Yanli Li, and Nicola Marzari.
\newblock First-principles simulation of electrochemical systems at fixed
  applied voltage: Vibrational stark effect for co on platinum electrodes.
\newblock 2009.
\newblock http://arxiv.org/abs/0901.0096v2.

\bibitem{carpar1}
C.~Cavazzoni and G.~L. Chiarotti.
\newblock {\em Comput. Phys. Commun.}, 123:56, 1999.

\bibitem{carpasqua0}
A.~Pasquarello, K.~Laasonen, R.~Car, C.~Lee, and D.~Vanderbilt.
\newblock {\em Phys. Rev. Lett.}, 69:1982, 1992.

\bibitem{sic}
M.~d'Avezac, M.~Calandra, and F.~Mauri.
\newblock {\em Phys. Rev. B}, 71:205210, 2005.

\bibitem{sit07}
Hoi-Land Sit, Matteo Cococcioni, and Nicola Marzari.
\newblock {\em J. Electroanal. Chem.}, 607:107--112, 2007.

\bibitem{sit06}
Hoi-Land Sit, Matteo Cococcioni, and Nicola Marzari.
\newblock {\em Phys. Rev. Lett.}, 97:028303, 2006.

\bibitem{ensemble}
N.~Marzari, D.~Vanderbilt, and M.~C. Payne.
\newblock {\em Phys. Rev. Lett.}, 79:1337, 1997.

\bibitem{cppre}
F.~Tassone, F.~Mauri, and R.~Car.
\newblock {\em Phys. Rev. B}, 50:10561, 1994.

\bibitem{Nose}
D.~J. Tobias, G.~J. Martyna, and M.~L. Klein.
\newblock {\em J. Phys. Chem.}, 97:12959, 1993.

\bibitem{martyna:2635}
Glenn~J. Martyna, Michael~L. Klein, and Mark Tuckerman.
\newblock {\em J. Chem. Phys.}, 97(4):2635--2643, 1992.

\bibitem{payne}
M.~C. Payne, M.~P. Teter, D.~C. Allen, T.~A. Arias, and J.~D. Joannopoulos.
\newblock {\em Rev. Mod. Phys.}, 64:1045, 1992.

\bibitem{marxhutter}
D.~Marx and J.~Hutter.
\newblock Modern methods and algorithms of quantum chemistry.
\newblock \uppercase{J}ohn von Neumann Institute for Computing, FZ J\"ulich,
  pp. 301--449 (2000).

\bibitem{Dubois}
V.~Dubois, P.~Umari, and A.~Pasquarello.
\newblock {\em Chem. Phys. Lett.}, 390:193, 2004.

\bibitem{Giustino}
F.~Giustino and A.~Pasquarello.
\newblock {\em Phys. Rev. Lett.}, 95:187402, 2005.

\bibitem{IRandRaman}
P.~Umari and A.~Pasquarello.
\newblock {\em Diam. Relat. Mater.}, 14:1255, 2005.

\bibitem{IRandRaman2}
P.~Umari and A.~Pasquarello.
\newblock {\em Phys. Rev. Lett.}, 95:137401, 2005.

\bibitem{IRandRaman3}
L.~Giacomazzi, P.~Umari, and A.~Pasquarello.
\newblock {\em Phys. Rev. Lett.}, 95:075505, 2005.

\bibitem{HyperRaman}
P.~Umari and A.~Pasquarello.
\newblock {\em Phys. Rev. Lett.}, 98:176402, 2007.

\bibitem{c60ph}
P.~Giannozzi and S.~Baroni.
\newblock {\em J. Chem. Phys.}, 100:8537, 1994.

\bibitem{dfpt3}
X.~Gonze.
\newblock {\em Phys. Rev. B}, 55:10337, 1997.

\bibitem{dfpt4}
X.~Gonze and C.~Lee.
\newblock {\em Phys. Rev. B}, 55:10355, 1997.

\bibitem{Dalcorso1}
A.~{Dal Corso}, A.~Pasquarello, and A.~Baldereschi.
\newblock {\em Phys. Rev. B}, 56:R11369, 1997.

\bibitem{Dalcorso1b}
A.~{Dal Corso}.
\newblock {\em Phys. Rev. B}, 64:235118, 2001.

\bibitem{dfptgga1}
F.~Favot and A.~{Dal Corso}.
\newblock {\em Phys. Rev. B}, 60:11427, 1999.

\bibitem{dfptgga2}
A.~{Dal Corso} and S.~de~Gironcoli.
\newblock {\em Phys. Rev. B}, 62:273, 2000.

\bibitem{Dalcorso_paw}
A.~{Dal Corso}.
\newblock {\em submitted}, 2009.

\bibitem{Dalcorso_so}
A.~{Dal Corso}.
\newblock {\em Phys. Rev. B}, 76:054308, 2007.

\bibitem{giustino07}
Feliciano Giustino, Marvin~L. Cohen, and Steven~G. Louie.
\newblock {\em Phys. Rev. B}, 76:165108, 2007.

\bibitem{anharm}
M.~Lazzeri and S.~de~Gironcoli.
\newblock {\em Phys. Rev. B}, 65:245402, 2002.

\bibitem{raman1}
M.~Lazzeri and F.~Mauri.
\newblock {\em Phys. Rev. Lett.}, 90:036401, 2003.

\bibitem{raman2}
M.~Lazzeri and F.~Mauri.
\newblock {\em Phys. Rev. B}, 68:161101(R), 2003.

\bibitem{KH}
D.~D. Koelling and B.~N. Harmon.
\newblock {\em J. Phys. C: Solid State Phys.}, 10(16):3107--3114, 1977.

\bibitem{RDFT}
A.~H. MacDonald and S.~H. Vosko.
\newblock {\em J. Phys. C: Solid State Phys.}, 12(15):2977--2990, 1979.

\bibitem{RDFT2}
A.~K. Rajagopal and J.~Callaway.
\newblock {\em Phys. Rev. B}, 7(5):1912--1919, Mar 1973.

\bibitem{TM}
N.~Troullier and Jos\'e~Lu\`{\i}s Martins.
\newblock {\em Phys. Rev. B}, 43(3):1993--2006, Jan 1991.

\bibitem{RRKJ}
Andrew~M. Rappe, Karin~M. Rabe, Efthimios Kaxiras, and J.~D. Joannopoulos.
\newblock {\em Phys. Rev. B}, 41(2):1227--1230, Jan 1990.

\bibitem{Kresse_pseudi}
G.~Kresse and J.~Hafner.
\newblock {\em J. Phys.-Condens. Mat.}, 6:8245, 1994.

\bibitem{ball2}
A.~Smogunov, A.~{Dal Corso}, and E.~Tosatti.
\newblock {\em Phys. Rev. B}, 70:045417, 2004.

\bibitem{ball3}
A.~{Dal Corso}, A.~Smogunov, and E.~Tosatti.
\newblock {\em Phys. Rev. B}, 74:045429, 2006.

\bibitem{Profeta_2003_a}
M.~Profeta, F.~Mauri, and C.~J. Pickard.
\newblock {\em J. Am. Chem. Soc.}, 125:541, 2003.

\bibitem{vandeWalle_1993_a}
C.~G. van~de Walle and P.~E. Bl\"ochl.
\newblock {\em Phys. Rev.}, B 47:4244, 1993.

\bibitem{Mauri_1996_a}
F.~Mauri and S.~G. Louie.
\newblock {\em Phys. Rev. Lett.}, 76:4246, 1996.

\bibitem{Mauri_1996_b}
F.~Mauri, B.~Pfrommer, and S.~G. Louie.
\newblock {\em Phys. Rev. Lett.}, 77:5300, 1996.

\bibitem{Pickard_2002_a_gtensor}
C.~J. Pickard and F.~Mauri.
\newblock {\em Phys. Rev. Lett.}, 88:086403, 2002.

\bibitem{converseNMR}
Timo Thonhauser, Davide Ceresoli, Arash Mostofi, Nicola Marzari, Raffaele
  Resta, and David Vanderbilt.
\newblock Electrostatics in periodic boundary conditions and real-space
  corrections.
\newblock {\em submitted to Phys. Rev. Lett.}, 2009.
\newblock http://arxiv.org/abs/0709.4429v1.

\bibitem{GougoussisUS}
C.~Gougoussis, M.~Calandra, A.~Seitsonen, and F.~Mauri.
\newblock {\em Phys. Rev. B, in press}, 2009.
\newblock http://arxiv.org/abs/0906.0897.

\bibitem{Gougoussis}
C.~Gougoussis, M.~Calandra, A.~Seitsonen, Ch. Brouder, A.~Shukla, and F.~Mauri.
\newblock {\em Phys. Rev. B}, 79:045118, 2009.

\bibitem{wannier90-url}
http://www.wannier.org/.

\bibitem{pierino98}
P.L. Silvestrelli, N.~Marzari, D.~Vanderbilt, and M.~Parrinello.
\newblock {\em Solid State Commun.}, 107:7, 1998.

\bibitem{calzolari04}
A.~Calzolari, I.~Souza, N.~Marzari, and M.~Buongiorno Nardelli.
\newblock {\em Phys. Rev. B}, 69:035108, 2004.

\bibitem{yslee05}
Young-Su Lee, M.~Buongiorno Nardelli, and Nicola Marzari.
\newblock {\em Phys. Rev. Lett.}, 95:076804, 2005.

\bibitem{yates07}
J.~R. Yates, X.~Wang, D.~Vanderbilt, and I.~Souza.
\newblock {\em Phys. Rev. B}, 75:195121, 2007.

\bibitem{xcrysden}
A.~Kokalj.
\newblock Xcrysden-a new program for displaying crystalline structures and
  electron densities.
\newblock {\em J. Mol. Graph. Model.}, 17:176--179, 1999.
\newblock http://www.xcrysden.org/.

\bibitem{Hump96}
William Humphrey, Andrew Dalke, and Klaus Schulten.
\newblock {VMD} -- {V}isual {M}olecular {D}ynamics.
\newblock {\em J. Mol. Graph. Model.}, 14:33--38, 1996.

\bibitem{casino}
http://www.tcm.phy.cam.ac.uk/~mdt26/casino2.html.

\bibitem{dmft}
D.~Korotin, A.~V. Kozhevnikov, S.~L. Skornyakov, I.~Leonov, N.~Binggeli, V.~I.
  Anisimov, and G.~Trimarchi.
\newblock {\em Eur. Phys. J. B}, 65:91--98, 2008.
\newblock http://dmft@qe-forge.org.

\bibitem{sax}
L.~Martin-Samos and G.~Bussi.
\newblock Sax: An open source package for electronic-structure and
  optical-properties calculations in the gw approximation.
\newblock {\em Comp. Phys. Comm., in press}, 2009.
\newblock http://sax-project.org.

\bibitem{dp}
http://dp-code.org.

\bibitem{exc}
http://www.bethe-salpeter.org.

\bibitem{guib}
A.~Kokalj.
\newblock Computer graphics and graphical user interfaces as tools in
  simulations of matter at the atomic scale.
\newblock {\em Comp. Mater. Sci.}, 28:155, 2003.
\newblock http://www-k3.ijs.si/kokalj/guib/.

\bibitem{rimini08}
P.~Giannozzi and C.~Cavazzoni.
\newblock Large-scale computing with quantum-espresso.
\newblock {\em Nuovo Cimento C, in press}, 32, 2009.

\bibitem{tg}
Juerg Hutter and Alessandro Curioni.
\newblock Car-parrinello molecular dynamics on massively parallel computers.
\newblock {\em ChemPhysChem}, 6:1788, 2005.

\bibitem{scalapack}
L.~S. Blackford, J.~Choi, A.~Cleary, J.~Demmel, I.~Dhillon, J.~Dongarra,
  S.~Hammarling, G.~Henry, A.~Petitet, K.~Stanley, D.~Walker, and R.~C. Whaley.
\newblock Scalapack: A portable linear algebra library for distributed memory
  computers - design issues and performance.
\newblock {\em SC Conference}, 0:5, 1996.

\bibitem{qbox}
Francois Gygi, Erik~W. Draeger, Martin Schulz, Bronis~R. de~Supinski, John~A.
  Gunnels, Vernon Austel, James~C. Sexton, Franz Franchetti, Stefan Kral,
  Christoph~W. Ueberhuber, and Juergen Lorenz.
\newblock {\em IBM J. Res. Dev.}, 52(1/2):137, 2008.

\bibitem{OpenMP}
{\em Using OpenMP:Portable Shared Memory Parallel Programming}.
\newblock The MIT Press, 2007.

\bibitem{RG:84}
Erich Runge and E.~K.~U. Gross.
\newblock {\em Phys. Rev. Lett.}, 52(12):997--1000, 1984.

\bibitem{TDDFT-Book}
M.~A.~L. Marques, C.~L. Ullrich, F.~Nogueira, A.~Rubio, K.~Burke, and E.~K.~U.
  Gross, editors.
\newblock {\em Time-Dependent Density Functional Theory}, volume 706 of {\em
  Lecture notes in Physics}.
\newblock Springer-Verlag, Berlin, Heidelberg, 2006.
\newblock {DOI}-10.1007/3-540-35426-3-17.

\bibitem{Onida:02}
Giovanni Onida, Lucia Reining, and Angel Rubio.
\newblock {\em Rev. Mod. Phys.}, 74(2):601--659, 2002.

\bibitem{tddfpt}
D.~Rocca, R.~Gebauer, Y.~Saad, and S.~Baroni.
\newblock {\em J. Chem. Phys.}, 128:154105, 2008.

\bibitem{Umari:arXiv0811.1453}
P.~Umari, Geoffrey Stenuit, and Stefano Baroni.
\newblock {\em Phys. Rev. B}, 79:201104(R), 2009.

\bibitem{sdg-vdw1}
H.-V. Nguyen and S.~de~Gironcoli.
\newblock {\em Phys. Rev. B}, 79:205114, 2009.

\bibitem{sdg-vdw2}
H.-V. Nguyen and S.~de~Gironcoli.
\newblock {\em Phys. Rev. B}, 79:115105, 2009.

\bibitem{viet-PhD}
H.-V. Nguyen.
\newblock PhD thesis, SISSA, 2008.
\newblock http://www.sissa.it/cm/thesis/2008/VietHuyNguyen\_PhDthesis.pdf.

\bibitem{yambo}
Andrea Marini, Conor Hogan, Myrta Gr{\"u}ning, and Daniele Varsano.
\newblock Yambo: an ab initio tool for excited state calculations.
\newblock {\em Comp. Phys. Commun., in press}, 2009.
\newblock http://www.yambo-code.org.

\bibitem{want}
A.~Calzolari, I.~Souza, N.~Marzari, and M.~Buongiorno Nardelli.
\newblock {\em Phys. Rev. B}, 69:035108, 2004.
\newblock http://www.wannier-transport.org.

\bibitem{qha}
http://qha@qe-forge.org.

\bibitem{pwtk}
A. Kokalj, \texttt{pwtk}: a Tcl scripting interface to PWscf,
  http://pwtk.qe-forge.org/.

\bibitem{vlab1}
P.~da~Silveira, C.~R.~S. da~Silva, and R.~M. Wentzcovitch.
\newblock {\em Comput. Phys. Commun.}, 178:186, 2008.

\bibitem{vlab2}
Cesar R.~S. da~Silva, P.~R.~C. da~Silveira, {B. B. Karki}, R.~M Wentzcovitch,
  P.~A. Jensen, E.~F. Bollig, M.~Pierce, G.~Erlebacher, and D.~A. Yuen.
\newblock {\em Phys. Earth Planet. Int.}, 163:321, 2007.
\newblock Special Issue: Computational Challenges.

\bibitem{vlab3}
http://dasilveira.cems.umn.edu:8080/gridsphere/gridsphere,
  http://www.vlab.msi.umn.edu.

\bibitem{vlab4}
R.~M. Wentzcovitch, B.~B. Karki, M.~Cococcioni, and S.~de~Gironcoli.
\newblock {\em Phys. Rev. Lett.}, 92:018501, 2004.

\bibitem{TFmixing}
D.~Raczkowski, A.~Canning, and L.~W. Wang.
\newblock {\em Phys. Rev. B}, 64:R121101, 2001.

\bibitem{arias}
T.~A. Arias, M.~C. Payne, and J.~D. Joannopoulos.
\newblock {\em Phys. Rev. B}, 45:1538, 1992.

\bibitem{cgdiago}
A.~Qteish.
\newblock {\em Phys. Rev. B}, 52:14497--14504, 1995.

\bibitem{mead}
C.~Alden Mead.
\newblock {\em Rev. Mod. Phys.}, 64:51, 1992.

\bibitem{grid}
R.~{di Meo}, A.~{Dal Corso}, P.~Giannozzi, and S.~Cozzini.
\newblock Calculation of phonon dispersions on the grid using quantum espresso.
\newblock In {\em Proceedings of the COST School, Trieste}, 2009.
\newblock to be published.

\bibitem{hf1}
F.~Gygi and A.~Baldereschi.
\newblock {\em Phys. Rev. B}, 34:4405, 1986.

\end{thebibliography}

\end{document}